\documentclass[aps,prl,notitlepage,nofootinbib,superscriptaddress,twocolumn,groupedaddress,showpacs]{revtex4}

\usepackage{amsmath}

\usepackage{amssymb}
\usepackage{graphicx,color}

\usepackage{amsmath,amsthm}
\usepackage{txfonts}

\usepackage{mathrsfs}

\usepackage{bm}

\newcommand{\Ref}[1]{(\ref{#1})}

\newcommand{\Z}{\mathbb{Z}}
\newcommand{\R}{\mathbb{R}}
\newcommand{\C}{\mathbb{C}}

\newcommand{\half}{\frac{1}{2}}

\newcommand{\ccirc}{\kern0.2ex\vcenter{\hbox{$\scriptstyle\circ$}}\kern0.2ex}

\newcommand{\Slc}{\mathrm{SL}(2,\mathbb{C})}

\newcommand{\Su}{\mathrm{SU}(2)}

\def\be{\begin{eqnarray}}
\def\ee{\end{eqnarray}}

\newcommand{\ch}{\mathcal H}

\newcommand{\ck}{\mathcal K}

\newcommand{\cs}{\mathcal S}

\newcommand{\sm}{\mathscr{M}}

  \newcommand{\Fl}{\mathfrak{L}}

\renewcommand{\a}{\alpha}
\renewcommand{\b}{\beta}
\newcommand{\g}{\gamma}

\newcommand{\eps}{\varepsilon}

\newcommand{\sig}{\sigma}

\renewcommand{\l}{\lambda}

\newcommand{\rmd}{\mathrm d}

\newcommand{\lt}{\left}
\newcommand{\rt}{\right}

\newcommand{\lag}{\left\langle}
\newcommand{\rag}{\right\rangle}

\newcommand{\Ar}{\mathbf{a}}

\newcommand{\background}{\lt(\mathring{J}_{f},\mathring{g}_{ve},\mathring{z}_{vf}\rt)}

\newcommand{\sn}{\mathscr{N}}

\newcommand{\re}{\mathrm{Re}}

\begin{document}

\sloppy

\title{\bf Einstein Equation from Covariant Loop Quantum Gravity in Semiclassical Continuum Limit}

\author{Muxin Han}
\affiliation{Department of Physics, Florida Atlantic University, 777 Glades Road, Boca Raton, FL 33431, USA}
\affiliation{Institut f\"ur Quantengravitation, Universit\"at Erlangen-N\"urnberg, Staudtstr. 7/B2, 91058 Erlangen, Germany}


\begin{abstract}

In this paper we explain how 4-dimensional general relativity and in particular, the Einstein equation, emerge from the spinfoam amplitude in loop quantum gravity. We propose a new limit which couples both the semiclassical limit and continuum limit of spinfoam amplitudes. The continuum Einstein equation emerges in this limit. Solutions of Einstein equation can be approached by dominant configurations in spinfoam amplitudes. A running scale is naturally associated to the sequence of refined triangulations. The continuum limit corresponds to the infrared limit of the running scale. An important ingredient in the derivation is a regularization for the sum over spins, which is necessary for the semiclassical continuum limit. We also explain in this paper the role played by the so-called flatness in spinfoam formulation, and how to take advantage of it.


\end{abstract}


\maketitle

\section{Introduction}\label{intro}

Loop quantum gravity (LQG) is an attempt toward the nonperturbative and background independent quantum theory of gravity \cite{book,review,review1}. The covariant approach of LQG is known as the spinfoam formulation \cite{rovelli2014covariant,Perez2012}, in which the quantum spacetime is understood by the spinfoam amplitude describing the transition between quantum spatial geometries. 

This paper focuses on the semiclassical behavior of the covariant LQG. A consistent quantum theory of gravity must reproduce general relativity (GR) as its semiclassical limit. In this paper, we explain how GR and the Einstein equation emerge from the covariant LQG. 

The analysis and results in this paper evolves from the recent extensive studies of spinfoam asymptotics (briefly reviewed in Section \ref{1}, see also e.g. \cite{CFsemiclassical,semiclassical,HZ,HHKR,Kaminski:2017eew}). It has been shown that if one doesn't consider the spin-sum, but consider the spinfoam (partial) amplitude with fixed spins, the large spin asymptotics of the amplitude give the Regge action of gravity, being a discretization of the Einstein-Hilbert action on the triangulation. 

However, the discussion on carrying out sum over spins and its semiclassical limit has been not sufficient in the literature, whose reason is explained in a moment. There has been a proposal of carrying out spin sum semiclassically in asymptotically large spins while sending Barbero-Immirzi parameter $\g$ to zero at the same time \cite{claudio1}. This proposal produces Regge equation (equation of motion from Regge action) from spinfoam amplitude. The idea of this type of limit has also been used in the graviton propagator computation from spinfoams \cite{propagator3,propagator2,propagator1,Alesci:2009ys,3pt}. 

The present work considers the semiclassical behavior of the spinfoam amplitude with an arbitrarily fixed Barbero-Immirzi parameter, and takes into account the sum over spins. The semiclassical limit in this situation turns out to have more interesting consequences. The reason why this situation wasn't sufficiently studied has been the question about the \emph{flatness} in spinfoam amplitudes. It was observed in \cite{flatness,frankflat,Perini:2012nd} that when one takes into account the sum over spins and studies the semiclassical limit, the spinfoam amplitude is dominant by the flat Regge geometry with all deficit angles vanishing\footnote{More precisely, the dominant geometries there have deficit angles vanishing modulo $4\pi\Z$.}. There has been worry in the LQG community that the flatness might be the obstruction of spinfoam amplitude to have a consistent semiclassical limit. However it has been suggested in \cite{Han:2016fgh} that the flatness, if treated properly, is a good property of spinfoam amplitude, which makes spinfoams well-behaved near the classical curvature singularity. Moreover it has also been suggested in \cite{freideltalk,Perini:2012nd} that the flatness should relate to the continuum limit of spinfoams, since deficit angles of discrete geometries indeed approach to zero in the continuum limit. Namely the flatness means that for spinfoam amplitude, the semiclassical limit should be taken together with the continuum limit\footnote{\cite{freideltalk} mentioned this limit as an analog of the hydrodynamical limit.}. The last point of view is one of the motivations of the present work.   

The situation is similar to the subtlety of interchanging limits in mathematical physics. We have two limits involved here (1) deficit angles $\eps_f\to0$ and (2) the refinement limit of triangulations. $\eps_f\to0$ relates to the lattice spacing $\ell \to0$ in Regge geometries since $\eps_f\sim \ell^2/\rho^2$ where $\rho$ is the curvature scale of the geometry approximated by Regge geometries \cite{FFLR}. If one takes firstly the limit (1) then takes the limit (2), one only obtains the flat geometry on the continuum. However if both limits are coupled and taken at the same time, instead of one after the other, we can recover arbitrary curved geometry by the limit \cite{BARRETT1994107}. In the derivation of the flatness \cite{flatness,frankflat,Perini:2012nd}, the treatment of spin sum effectively leads to $\eps_f\to0$ on a fixed triangulation (before the refinement limit). In order to implement the proper limit, taking (1) and (2) at the same time, the spin sum has to be treated differently, which should open a window of small but nonvanishing $\eps_f$, to let $\eps_f\to0$ couple nontrivially to the refinement limit. 

The desired window can be given by the treatment in \cite{Han:2016fgh}, where a damping factor is inserted in the sum over spins. The damping factor regularizes the spin sum by suppressing the contribution from spins far away from a given spin configuration $J_0$. The damping is turned off together with the large $J_0$ limit. The regularization procedure indeed produces a small window of nonvanishing deficit angle. Then the authors are able to show that the effective action at $J_0$ from spinfoam amplitude approximate the Einstein-Hilbert action, when $J_0$ corresponds to a set of geometrical triangle areas on the triangulation. 

In this paper we propose an improved regularization scheme in Section \ref{2Reg}, which is more suitable in analyzing the sum over contributions from different spin configurations. It is based on the following observations: The spinfoam asymptotics (with fixed spins) reproduce Regge geometries and the Regge action when the fixed spins are \emph{Regge-like}, i.e. the spins $\vec{J}(\ell)$ which can be expressed as triangle areas in terms of a set of edge lengths $\{\ell\}$ on the triangulation (the spins only need to be close to Regge-like in order to produce the Regge geometry and the Regge action). Regge-like spins locate in a submanifold $\sm_{Regge}$ in the space of all spin configurations. Motivated by this property, we decompose the sum over spins in the spinfoam amplitude into a sum over Regge-like spins along $\sm_{Regge}$ and a sum along transverse directions which contains non-Regge-like spins. As an equivalent way to understand the flatness, its origin is the fact that non-Regge-like spins in transverse directions contribute nontrivially to the amplitude in the large spin asymptotics. Based on the above observations, we propose to only regularize the spin sum in transverse directions instead of the regularization in all directions as in \cite{Han:2016fgh}. The regularization is made by inserting a Gaussian distribution with width $\delta^{-1/2}$ in the transverse spin sum. The Gaussian produces the damping at the infinity in transverse directions. The regulator will be removed by $\delta\to0$ in the end together with the continuum limit.

The regularized sum in transverse directions can be computed explicitly, which produces a Gaussian of width $\delta^{1/2}$ peaked at a submanifold in the space of spinfoam variables. After carrying out the transverse spin sum, we are only left with the sum over Regge-like spins. Schematically the spinfoam amplitude reduces to be the following type
\be
Z=\sum_{{J}(\ell)}\int\rmd\mu(X)\, e^{S\lt[J(\ell),X\rt]}\,D_\delta(\ell,X)\label{Z0}
\ee
where $X$ labels spinfoam variables in addition to spins in the integral representation of $Z$. $S$ is the spinfoam action used in the asymptotical analysis. $D_\delta$ contains the Gaussian of width $\delta^{1/2}$ mentioned above.

The action $S$ in Eq.\Ref{Z0} only involves Regge like spins. So the results of large spin asymptotics can be immediately applied to the semiclassical analysis in Section \ref{RESDA}. We consider the spinfoam state sum in the semiclassical regime. Namely we focus on a neighborhood $\sn_{Regge}\subset\sm_{Regge}$ such that the spins within $\sn_{Regge}$ are uniformly large. We introduce a parameter $\l\gg1$ as a typical value of spin in $\sn_{Regge}$. The spin sum in Eq.\Ref{Z0} is performed in $\sn_{Regge}$. Then entire domain of the spin sum including transverse directions is denoted by $\sn$. The spinfoam amplitude is denoted by $Z_{\sn,\delta}(\ck)$ depending on 3 types of parameters: the spin sum domain $\sn$ of large spins $J\sim\l$, the regulator $\delta$, and the triangulation $\ck$. An interesting regime where $Z_{\sn,\delta}(\ck)$ exhibits desired semiclassical behavior is 
\be
\l\gg\delta^{-1}\gg1\label{gggg0}
\ee
In this regime, $Z_{\sn,\delta}(\ck)$ is dominated by the critical points of $S\lt[J(\ell),X\rt]$, which has been extensively studied in the literature \cite{CFsemiclassical,semiclassical,HZ,HZ1,hanPI}). With respect to $\int\rmd\mu(X)$, the critical points give Regge geometries on $\ck$. Taking into account $\sum_{{J}(\ell)}$ reduce the critical points to the ones corresponding to geometries satisfying the Regge equation (the equation of motion of the Regge action). Because of Eq.\Ref{gggg0}, the leading contributions are computed by evaluating $D_\delta$ at the critical points. Then the Gaussian in $D_\delta$ together with the Regge equation constrains the deficit angles $\eps_f$ to be small (but nonvanishing)
\be
|\g\eps_f|\leq\delta^{1/2}.\label{smalleps0}
\ee
$\g$ is an fixed $O(1)$ parameter throughout our discussion. Note that there exists some discrete ambiguities of the above constraint, due to the periodicity of the integrand in Eq.\Ref{Z0}. But the ambiguities can be removed by suitably choosing $\sn_{Regge}$. The regime where the Regge equation and the constraint Eq.\Ref{smalleps0} emerge from the spinfoam amplitude is referred to as the \emph{Einstein-Regge (ER) regime} in Section \ref{4ERR}. 

As promised, the regularization of the spin sum opens a small window for nontrivial $\eps_f$. Small $\eps_f$ relates to the continuum limit of Regge geometries, because $|\eps_f|\sim {\ell^2}/{\rho^2}$ \cite{FFLR} where $\rho$ is the typical curvature radius of the smooth geometry approximated by the Regge geometry. $|\eps_f|\ll 1$ relates to $\ell\ll\rho$. $\delta$ behaves as the bound of error in approximating smooth geometries by Regge geometries. The emerging smooth geometries have nontrivial curvatures.

In the ER regime, the configurations contributing dominantly the spinfoam amplitude contains the Regge geometries satisfying Regge equation, and approximating (curved) smooth geometries. Regge geometries failing to approximate any smooth geometry are suppressed by the amplitude.

Eq.\Ref{smalleps0} indicates that the regulator $\delta$ relates to the continuum limit. The window of nontrivial $\eps_f$ allows us to couple $\eps_f\to0$ to the refinement limit of the triangulation. The continuum limit at the semiclassical level is discussed in Section \ref{SCL}. We consider an infinite sequence of triangulations given by the refinement, such that all vertices of triangulations form a dense set in the 4-manifold where triangulations are embedded. A sequence of spinfoam amplitudes $Z_{\sn,\delta}(\ck)$ are defined on the sequence of triangulations. We let the limit $\delta\to0$ couple to the refinement, i.e. $\delta\to0$ is taken together with the continuum limit.  

On the other hand, the typical spin value $\l$ has to increase in refining the triangulation. Refining the triangulation increases the number of degrees of freedom in spinfoam amplitude. It then requires a larger $\l$ to suppress the quantum correction, so that the semiclassical behavior stands out as the leading order (see Section \ref{SCL}).

The semiclassical continuum limit involves taking simultaneously 3 limits: triangulation refinement limit, $\l\to\infty$, and $\delta\to 0$. The limits are implement to the sequence of $Z_{\sn,\delta}(\ck)$. At each $Z_{\sn,\delta}(\ck)$ in the sequence, Eq.\Ref{gggg0} has to be satisfied, in order to keep a nontrivial ER regime. As a result, we obtain sequences of Regge geometries approaching smooth geometries in the limit. Each Regge geometry in each sequence (a) satisfies Regge equation, (b) satisfies small deficit angle constraint Eq.\Ref{smalleps0}, and (c) contributes dominantly to the corresponding $Z_{\sn,\delta}(\ck)$. We are able to achieve all (a), (b), and (c) because each Regge geometry in each sequence is inside the ER regime of the corresponding $Z_{\sn,\delta}(\ck)$.  

At first sight, $\l\to\infty$ might seem contradicting to the continuum limit, by the LQG relation $\Ar=\g\l\ell_P^2$ for the triangle areas. There is no contradiction because $\Ar$ is a dimensionful quantity, and the continuum limit corresponds to zoom out to larger length unit, such that the numerical value of $\ell_P^2$ measured by the unit shrinks in a faster rate than $\l\to\infty$. This observation motivates us to associates each triangulation and $Z_{\sn,\delta}(\ck)$ a mass scale $\mu$ whose $\mu^{-1}$ is a length unit. The refinement limit is labelled by the infrared (IR) limit $\mu\to0$. All parameters of $Z_{\sn,\delta}(\ck)$ have nontrivial running with $\mu$, i.e. 
\be
\ck=\ck_\mu,\ \l=\l(\mu),\ \delta=\delta(\mu),\ Z_{\sn,\delta}(\ck)=Z_{\sn(\mu),\delta(\mu)}(\ck_\mu)
\ee
Here $\l(\mu)$ increase monotonically as $\mu\to0$ while $\delta(\mu)$ decrease monotonically. Eq.\Ref{gggg0} is satisfied at each $\mu$. The dependence of $\l$ on $\mu$ displays that the semiclassical limit is coupled to the continuum limit. Given the running scale $\mu$, on each $\ck_\mu$, the area is expressed as
\be
\Ar(\mu)=\g\l(\mu)\ell_P^2=a(\mu)\mu^{-2}\label{amu}
\ee
The area in the $\mu^{-2}$ unit, $a(\mu)$, shrinks and approaches to zero in the IR limit $\mu\to0$. In Regge geometries, the value of typical edge length $a(\mu)^{1/2}$ in the $\mu^{-1}$ unit approaches to zero as the refinement limit, which orders the sequence of Regge geometries to approach the smooth geometry at IR. Smooth geometries living at IR are associated with the largest length unit $\mu^{-1}\to\infty$.

The above discussion exhibits how scales and a renormalization-group-like behavior emerge from the spinfoam formulation which originally is scale independent. Possible ways of associating scales $\mu$ to triangulations $\ck_\mu$ are classified in Section \ref{running}.

We have obtained from the spinfoam amplitude sequences of Regge geometries solving Regge equations, which converge to smooth geometries in the semiclassical continuum limit. Generically the resulting smooth geometries are solutions of the continuum Einstein equation. Although the general mathematical proof for the convergence of Regge solutions to Einstein equation solutions is not available in the literature, extensive studies of the Regge calculus provide many analytical and numerical results, which all support the convergence, and demonstrate the Regge calculus as a useful tool in numerical relativity (see e.g. \cite{Williams:1991cd,Gentle:2002ux} for reviews). Among the results, there has been a rigorous proof of the convergence in the linearized Regge calculus and linearized Einstein equation \cite{0264-9381-5-12-007,0264-9381-5-9-004,Christiansen2011}. Results in the nonlinear regime include e.g. Kasner universe, Brill waves, binary black holes, FLRW universe etc \cite{Gentle:2012tc,Gentle:1997df,Gentle:2002ux,Gentle:1998qg,Liu:2015gpa}. There has also been the convergence result by certain average of Regge equations \cite{Miller:1995gz}.

A key observation in all convergence results is that the deviation of Regge calculus from general relativity is essentially the non-commutativity of rotations in the discrete theory, while the error from the non-commutativity is of higher order in edge lengths \cite{Gentle:2008fy} \footnote{The author thanks Warner Miller for pointing this out.}. 

We conclude that for any sequence of Regge solutions converging to the solution of Einstein equation, the Regge solutions can be produced from the sequence of spinfoam amplitudes $Z_{\sn(\mu),\delta(\mu)}(\ck_\mu)$ as dominant configurations in the semiclassical approximation. The solution of the continuum Einstein equation lives at the IR limit $\mu\to 0$. The convergence to gravitational waves of the linearized Einstein equation in \cite{0264-9381-5-12-007} leads to a mathematically rigorous example for the emergence of Einstein equation from the spinfoam amplitude. 


There is a different argument for the emergence of Einstein equation from the spinfoam amplitude, by the convergence of effective actions (see Section \ref{SCL}). The analysis in this paper proposes a different regularization scheme from the one in \cite{Han:2016fgh}. However the results of the effective action in \cite{Han:2016fgh} and \cite{lowE,LowE1,lowE2} can be reproduced here. The effective action relates to $S[J(\ell),X]$ in Eq.\Ref{Z0} evaluated at critical points of $\int\rmd\mu(X)$ as $\l\gg1$ (before carrying out $\sum_{J(\ell)}$). $S[J(\ell),X]$ at critical points gives Regge actions evaluated at Regge geometries with small $\eps_f$ by Eq.\Ref{smalleps0}. When we consider the sequence $Z_{\sn(\mu),\delta(\mu)}(\ck_\mu)$ and take the semiclassical continuum limit. Regge actions converge to the Einstein-Hilbert action on the continuum, when Regge geometries converge to the smooth geometry \cite{cheeger1984,BARRETT1994107}. Translating the known convergence result to our context uses the length unit $\mu^{-1}$. We apply Eq.\Ref{amu} to the Regge action $\frac{1}{\ell_P^2}\sum_{f}\Ar_f(\mu)\eps_f(\mu)$ from $S[J(\ell),X]$ in $Z_{\sn(\mu),\delta(\mu)}(\ck_\mu)$ \footnote{$\Ar_f(\mu)=\g J_f(\mu)\ell_P^2=a_f(\mu)\mu^{-2}$.}:
\be
\frac{1}{\mu^2\ell_P^2}\sum_{f}a_f(\mu)\eps_f(\mu)\to \frac{1}{\mu^2\ell_P^2}\int\rmd^4x\sqrt{-g}\, R \label{EH0}
\ee
where the convergence happens as the edge length $a(\mu)^{1/2}\to 0$ at IR \footnote{The convergence requires the fatness of simplices to be bounded away from zero in addition to shrinking edge lengths, see \cite{cheeger1984,BARRETT1994107} for details.}. Smooth geometries and $\int\rmd^4x\sqrt{-g}\, R $ live in the IR limit $\mu\to0$. $\sum_{J(\ell)}$ (or $\sum_\ell$) in Eq.\Ref{Z0} sums all convergence sequences of Regge geometries, thus equivalently sums all smooth geometries in the limit. The spinfoam amplitude becomes a functional integral of Einstein-Hilbert action in the continuum (see Section \ref{SCL} for details). Then $\mu\to 0$ in Eq.\Ref{EH0} leads to the continuum vacuum Einstein equation
\be
R_{\mu\nu}=0
\ee
by the variational principle.

The quantum behavior of spinfoams near a classical curvature singularity derived in \cite{Han:2016fgh} can be reproduced in the present regularization scheme. Large-$J$ and Eq.\Ref{smalleps0} show that the semiclassical approximation is valid only in the regime that ($\ell^2\sim\Ar_f$)
\be
\ell_P\ll\ell\ll\rho.\label{llll0}
\ee
However a large curvature may violate $\ell_P\ll\rho$, and lead to the incompatibility between $\ell\ll\rho$ and large-$J$. Therefore the semiclassical analysis in this paper is not valid near the curvature singularity. Similar to \cite{Han:2016fgh}, spinfoams near the singularity are of small spins, in order that the amplitudes are not suppressed. It shows that the classical singularity corresponds to the quantum regime of spinfoams, where the theory is well defined but with large quantum fluctuations. 

As a key ingredient in the argument, Eq.\Ref{smalleps0} comes from the regularized flatness. It shows that the flatness is a good property of the spinfoam amplitude, which guarantees spinfoams behave correctly near a classical singularity.

We remark that the presentation in this paper uses the spinfoam models of Engle-Pereira-Rovelli-Livine/Freidel-Krasnov (EPRL/FK), both in Lorentzian and Euclidean signatures \cite{EPRL,FK}. But the discussion and results are valid or any other spinfoam models which have both the correct large spin asymptotics, and the flatness (e.g. the model with time-like tetrahedra \cite{Conrady:2010kc} and its recent asymptotical analysis \cite{Kaminski:2017eew}).

The architecture of this paper is as follows: Section \ref{1} provides a review on the recent development of the spinfoam large spin asymptotics. Section \ref{2Reg} discusses the regularization of the spin sum along directions transverse to the submanifold $\sm_{Regge}$ of Regge-like spins. Section \ref{RESDA} analyzes the semiclassical approximation of the regularized spinfoam amplitude, which gives the Regge equation and small deficit angle constraint Eq.\Ref{smalleps0}. Section \ref{RESDA} defines the Einstein-Regge regime of the spinfoam amplitude, in which the amplitude exhibits the desired semiclassical property. Section \ref{SCL} discusses the semiclassical continuum limit of sequences of spinfoam amplitudes, which approaches the continuum Einstein equation. Section \ref{running} classifies possible runnings of scales $\mu$ associated to triangulations.





\section{Large-$J$ Asymptotics of Spinfoam Amplitude}\label{1}

We consider the EPRL/FK spinfoam amplitude $Z(\ck)$ defined on a triangulation $\ck$. $Z(\ck)$ has the following integral representation \cite{hanPI}. 
\be
&&Z(\mathcal{K})= \sum_{J_f} \prod_f \dim(J_f)A_{J_f}(\mathcal{K})\label{Z}\\
&=& \sum_{J_f}\prod_f \dim(J_f) \int_{\Slc} \prod_{(v,e)} \rmd g_{ve} \int_{\mathbb{CP}^1}\prod_{v\in\partial f} \rmd z_{vf}~ e^{S[J_f, g_{ve}, z_{vf}]}\nonumber
\ee
$v$, $e$ and $f$ label the 4-simplices, tetrahedra and triangles. They equivalently label the vertices, dual edges and faces in the dual complex $\mathcal{K}^*$.  $J_f\in \Z_+/2$ are SU(2) spins associated to triangles $f$. $g_{ve}\in \Slc$ are associated to half-edges $(v,e)$ in $\ck^*$ where $v$ is a end-point of $e$. $z_{vf}$ are 2-spinors modulo complex rescaling.  The spinfoam action $S[J_f, g_{ve}, z_{vf}]$ reads
\be
S[J_f, g_{ve}, z_{vf}]&=&\sum_{f} J_fF_f[g_{ve}, z_{vf}]\nonumber\\
F_f[g_{ve}, z_{vf}]&=& \ln \prod_{e\subset\partial f}\frac{\lag g_{ve}^{\dagger}z_{vf}, g_{v'e}^{\dagger}z_{v'f}\rag^2}{\lag g_{ve}^{\dagger}z_{vf}, g_{ve}^{\dagger}z_{vf}\rag\lag g_{v'e}^{\dagger}z_{v'f}, g_{v'e}^{\dagger}z_{v'f}\rag}\nonumber\\
&+& i\g \ln \prod_{e\subset\partial f} \frac{\lag g_{ve}^{\dagger}z_{vf}, g_{ve}^{\dagger}z_{vf}\rag}{\lag g_{v'e}^{\dagger}z_{v'f}, g_{v'e}^{\dagger}z_{v'f}\rag}.\label{S}
\ee
Here $\langle,\rangle$ is an SU(2) invariant Hermitian inner product between 2-spinors. $S$ is defined modulo $2\pi i\Z$ because of $J\in\Z/2$, while $F_f$ is defined modulo $4\pi i\Z$. The Barbero-Immirzi parameter $\gamma\in\R$ is treated as a constant of $O(1)$ in this paper. It is straightforward to show that the real part of $F_f$ is non-positive $\re F_f\leq 0$ by using Cauchy-Schwarz inequality \cite{hanPI}. 

$Z(\mathcal{K})$ is the spinfoam amplitude in Lorentzian signature. The amplitude in Euclidean signature is written in a similar manner. Differences from Eq.\Ref{Z} contains that integrals over $\Slc$ are replaced by integrals over $(g_{ve}^+,g_{ve}^-)\in\mathrm{SO}(4)$, and integrals over $z_{vf}$ are replaced by integrals over 2-spinors $\xi_{ef}$ (one for each pair $(e,f)$ with $e\subset f$ in $\ck^*$), where $\xi_{ef}$ is normalized by the Hermitian inner product on $\C^2$. $F_f$ for Euclidean amplitude reads \cite{CFsemiclassical,semiclassicalEu,HZ1}
\be
F_f[g^{\pm}_{ve}, \xi_{ef}]=\sum_{\pm}\sum_{v\in f}\frac{1\pm\g}{2}j_f\ln\lag\xi_{ef}\big|(g_{ve}^{\pm})^{-1}g_{ve'}^\pm\big|\xi_{e'f}\rag
\ee
The above presents the expression of the Euclidean amplitude with $\g<1$. The expression for $\g>1$ can be found in \cite{semiclassicalEu}.

In the following we often present the analysis in the notation of Lorentzian amplitude. The same analysis can be applied to Euclidean amplitude. The result is valid for both signatures. 

The asymptotical analysis of the partial amplitude $A_{J_f}(\mathcal{K})$ as $J_f$ uniformly large has been well-developed by the recent progress \cite{CFsemiclassical,semiclassical,HZ,HZ1,hanPI,LowE1}. Since $S$ is linear to $J_f$, as $J_f$ uniformly large, $A_{J_f}(\mathcal{K})$ is dominated by contributions from the critical points of the action $S[J_f, g_{ve}, z_{vf}]$, i.e. configurations $\background$ satisfying $\re S=0$ and $\partial_g S=\partial_z S=0 $. Importantly, the critical points can be interpreted as simplicial geometries (Regge geometries) on the 4d triangulation. The spins $\mathring{J}_f$ are interpreted as triangle areas $\mathring{\mathbf{a}}_f=\g \mathring{J}_f\ell_P^2$. When the triangulation is sufficiently refined, the critical points can approximate arbitrary geometries on a 4-dimensional manifold.

It is shown in \cite{hanPI,HZ} that at a critical point $\background$ corresponding to a nondegenerate Regge geometry with globally orientation and global time-orientation, its leading contribution to $A_{J_f}(\mathcal{K})$ gives the Regge action:
\be
A_{J_f}(\mathcal{K})\sim \exp\lt(\frac{i}{\ell_P^2} \sum_{f} \mathring{\mathbf{a}}_f \mathring{\eps}_f+\frac{i}{\ell_P^2} \sum_{f\subset\partial\ck} \mathring{\mathbf{a}}_f \mathring{\Theta}_f+\cdots\rt),\label{asymp}
\ee
where $\mathring{\eps}_f,\mathring{\Theta}_f$ are the bulk deficit angle and boundary dihedral angle from the geometrical interpretation of $\background$. The asymptotic formula of $A_{J_f}(\mathcal{K})$ is given by a sum over critical points weighted by the contribution from each critical point.  

Note that it is possible to have time non-oriented geometries from critical points. In this case, $\mathring{\eps}_f$ is replaced by $\mathring{\eps}_f\pm\g^{-1}\pi$ in Eq.\Ref{asymp}. See \cite{HZ} for details.

Eq.\Ref{asymp} holds for \emph{Regge-like} spins $J_f$. Namely, it requires spins $\mathring{J}_f$ can be expressed as areas in terms of edge-lengths $\ell$ from a Regge geometry on the triangulation. 
\be
\g \tilde{J}_f(\ell)=\frac{1}{4}\sqrt{2(\ell^2_{ij}\ell^2_{jk}+\ell^2_{ik}\ell^2_{jk}+\ell^2_{ij}\ell^2_{ik})-\ell_{ij}^4-\ell_{ik}^4-\ell_{jk}^4}.\label{area-length}
\ee
where $\ell$'s are the edge lengths (in Planck unit) of the triangle $f$. \emph{Regge-like} spins span a subspace in the space of all spins\footnote{In general for non-degenerate simplicial 4d manifolds the number of triangles is greater than the number of edges. }. 

The situation of non-Regge-like spins are subtle. Non-Regge-like spins $J_f$ doesn't lead to any solution to the critical equations $\re S=\partial S=0$. Especially $\re S<0$ for any solution to $\partial S=0$ \footnote{To study the asymptotics with non-Regge-like spins, the equation of motion should be replaced by $\partial \cs=0$ where $\cs$ is the analytic continuation of $S$. See \cite{lowE,LowE1} for detail.} with non-Regge-like $J_f$. Although critical equations are not satisfied, the contribution to spinfoam spin-sum are non-negligible \cite{LowE1,Han:2016fgh,frankflat}. Indeed, by the stationary phase approximation (see Theorem 7.7.5 and 7.7.1 in \cite{stationaryphase}), in case there is no critical point in the region of integral $\int_K e^{\l S}\rmd\mu$, 
\be
\lt|\int_K e^{\l S(x)}\rmd \mu(x)\rt|\leq C\lt(\frac{1}{\l}\rt)^k\sup_K\frac{1}{\lt(|S'|^2+ \mathrm{Re}(S)\rt)^{k}}
\ee
the integral decays faster than $(1/\l)^k$ for all $k\in\Z_+$, provided that $\sup([|S'|^2+\mathrm{Re}(S)]^{-k})$ is finite (i.e. doesn't cancel the $(1/\l)^k$ behavior in front). But for the non-Regge-like $J_f$, the corresponding $A_{J_f}(\ck)$ may not decay faster than $(1/\l)^k$ for all $k\in\Z_+$. It happens for non-Regge-like spins close to Regge-like $J_f=\l j_f$ ($\l\gg 1$) with the small gap $\Delta j_f\sim\frac{1}{2\l}$. In this case, $\sup([|S'|^2+\mathrm{Re}(S)]^{-k})$ is likely to be large and cancel the $(1/\l)^k$ behavior. Therefore the non-Regge-like spins have nontrivial contribute to the spinfoam spin sum. 



\section{Regularizing Non-Regge-like Spin Sum}\label{2Reg}

In order to understand the contribution from non-Regge-like spins, we split the spin-sum into a sum over Regge-like spins and a sum over non-Regge-like spins in the following analysis. Then the Non-Regge-like spin-sum is carried out explicitly, with a regulator inserted, while the Regge-like spin-sum is treated by the usual stationary phase approximation.

The space of internal spins $J_f$, $\Fl_J$, is a cubic lattice in the smooth space $\sm_J\simeq \R^{N_f}$ ($J_f$ at different $f$ can be regarded as independent in the spin sum, see Appendix \ref{free} for an explanation). We define the submanifold $\sm_{Regge}$ to be the image of the smooth embedding in Eq.\Ref{area-length} from the space of edge-lengths $\sm_\ell$ into $\sm_J$. We denote by $\tilde{J}_f(\ell)$ the image of the embedding from a given $\{\ell\}$. $\tilde{J}_f(\ell)$ is a smooth function defined by Eq.\Ref{area-length}, and may not be a half-integer.

Given a compact neighborhood $\sn_{Regge}$ in $\sm_{Regge}$ which contains $\tilde{J}_f(\ell)$ all satisfying $\tilde{J}_f(\ell)\gg 1$\footnote{$\sm_{Regge}$ may have self-intersections, but $\sn_{Regge}$ is always obtained as the smooth image of a neighborhood of $\ell$'s in the space of edge lengths. }, we define local coordinates $({\ell},\tilde{t})$ in $\sm_J$, where edge-lengths $\ell$ are coordinates in $\sm_{Regge}$, $\{\tilde{t}_i\}_{i=1}^M$ are transverse coordinates to $\sm_{Regge}$. We denote the coordinate basis for $\tilde{t}_i$ by $\hat{e}^i=((\hat{e}^i)_f)_f$, and choose $\sn$ to be the coordinate chart. $\hat{e}^i$ ($i=1,\cdots,M$) may be assumed as constant vectors in $\R^{N_f}$. So that the coordinate axises of $t_i$ are straight lines in $\R^{N_f}$. The transverse submanifolds coordinatized by $t_i$ are parallel planes $\R^M\hookrightarrow \R^{N_f}$. This assumption can always be achieved locally in a compact neighborhood $\sn_{Regge}$. The transverse plane located at $\{\ell\}$ is denoted by $\sm_{NR}(\ell)\simeq\R^M$.

For any set of internal spins $\vec{J}\in\sn$, it is expressed in the $({\ell},\tilde{t})$ coordinate, in which $\ell$'s give a unique $\vec{\tilde{J}}(\ell)\in\sn_{Regge}$. So $\vec{J}$ is written as
\be
\vec{J}=\vec{\tilde{J}}(\ell)+\sum_{i=1}^M \tilde{t}_{i}\hat{e}^i,\quad \text{with}\quad J_f(\ell)\gg1\label{vecJ}
\ee
Recall that $\vec{\tilde{J}}(\ell)$ are in general not spins. We define $\vec{J}(\ell)$ to be a set of spins in the transverse plane $\sm_{NR}(\ell)$, at the same $\{\ell\}$ as the ones determining $\vec{\tilde{J}}(\ell)$, and require $\vec{J}(\ell)$ has the shortest distance to $\vec{\tilde{J}}(\ell)$ measured in $\R^{N_f}$. $\vec{J}(\ell)$ defined in this way might not be unique. But when there are multiple choices, we make an arbitrary choice of $\vec{J}(\ell)$. The resulting $\vec{J}(\ell)$ is a representative of $\vec{\tilde{J}}(\ell)\in\sn_{Regge}$. Obviously the spins $\vec{J}$ can also be written as $\vec{J}=\vec{{J}}(\ell)+\sum_{i=1}^M \tilde{t}_{i}\hat{e}^i$ using the representative. Given that both $\vec{J},\vec{{J}}(\ell)$ are spins, then $\sum_{i=1}^M t_{i}\hat{e}^i$ are half-integers, so that $\vec{{J}}(\ell)+n\sum_{i=1}^M \tilde{t}_{i}\hat{e}^i$ are also spins when $n\in\Z$. Spins in $\sm_{NR}(\ell)$ form a $M$ dimensional periodic lattice $\Fl_{NR}(\ell)$, whose lattice basis is denoted by $\{\hat{e}^i(\ell)\}_{i=1}^M$. Therefore, any internal spins $\vec{J}\in\sn$ can be expressed as
\be
\vec{J}=\vec{J}(\ell)+\sum_{i=1}^M t_{i}\hat{e}^i(\ell),\quad \text{with}\quad J_f(\ell)\gg1\label{vecJ1}
\ee
where $t_i\in\Z$.

That $\Fl_{NR}(\ell)$ is a periodic lattice is equivalent to the existence of parallel $M$-dimensional lattice planes in $\Fl_J$ intersecting $\sn_{Regge}$ transversely, which is always true locally (See Appendix \ref{lattice} for an explanation). The local property is sufficient for the present discussion. 

$\vec{J}(\ell)$ in Eq.\Ref{vecJ1} is a representative of Regge-like spins, although it might not precisely located at $\sn_{Regge}$. Its distance to $\sn_{Regge}$ is at most of $O(1)$ \footnote{$\vec{J}(\ell)$ generically satisfy the triangle inequality everywhere on $\ck$ since $\vec{\tilde{J}}(\ell)$ do.}. The large-$J$ asymptotics of $A_{{J}(\ell)}$ is the same as the situation of Regge-like spins in Eq.\Ref{asymp} by the argument at the end of last section (see also \cite{LowE1}). Non-Regge-like spins with $t_i\neq 0$ in each $\Fl_{NR}(\ell)$ is going to be summed explicitly under certain regularization, before the stationary phase approximation.




If we denote by $\lag\ ,\ \rag$ the Euclidean inner product in $\R^{N_f}$, the spinfoam action is written as
\be
\sum_f J_f F_f\equiv \lag \vec{J},\vec{F}\rag=\lag \vec{J}(\ell),\vec{F}\rag+\sum_i t_i\lag \hat{e}^i(\ell),\vec{F}\rag.
\ee
We define the spinfoam state sum in the coordinate chart $\sn$ by restricting the spin-sum in $\sn$,
\be
\!\!\!\!\!&&Z_\sn(\ck)=\sum_{\vec{J}\in\sn}\prod_f\dim(J_f)\int \rmd g_{ve}\rmd z_{vf}\, e^{\lag \vec{J},\vec{F}\rag}\nonumber\\
&=&\!\!\!\!\sum_{\vec{J}(\ell)}\sum_{t_i\in\Z}\mu(\ell,t)\int \rmd g_{ve}\rmd z_{vf}\, e^{\lag \vec{J}(\ell),\vec{F}\rag+\sum_i t_i\lag \hat{e}^i(\ell),\vec{F}\rag}.\label{ZN}
\ee
where $\mu(\ell,t)\equiv2^{N_f}\prod_f\lt({J}_f(\ell)+\sum_{i=1}^M t_{i}(\hat{e}^i)_f(\ell)\rt)$. The spin-sum only involves spins in the bulk. Boundary spins are set to be Regge-like $J_f={J}_f(\ell)$, $f\in\partial\ck$, as the boundary condition.


We perform a regularization (or deformation) of $\sum_{t_i\in\Z}$ by inserting a Gaussian weight 
\be
\sum_{t_i\in\Z}\to \sum_{t_i\in\Z} e^{-\frac{\delta}{4}\sum_{i=1}^M{t}_i{t}_i},\label{deform1}
\ee
The regulators $\delta\ll1$, which will be turned off appropriately by $\delta \to 0$ in the end. The amplitude with the insertion $e^{-\frac{\delta}{4}\sum_{i=1}^M{t}_i{t}_i}$ is denoted by $Z_{\sn,\delta}(\ck)$, which is a deformation from the original amplitude. When $\delta\to0$, $Z_{\sn,\delta}(\ck)$ returns to the spinfoam amplitude restricted to the domain $\sn$ of spins. The deformation turns out to be crucial in opening a small window of nontrivial curvature. The exponentially damping behavior of $e^{-\frac{\delta}{4}\sum_{i=1}^M{t}_i{t}_i}$ at $t\to\infty$ also justifies the Poisson resummation in the following.

We treat the sum over $t_i$ via the Poisson resummation (see Appendix \ref{EMF} for some discussions about the sum): 
\be
&&\sum_{t_i\in\Z}\mu(\ell,t)\, e^{-\frac{\delta}{4}\sum_{i=1}^M{t}_i{t}_i+\sum_i t_i\lag \hat{e}^i(\ell),\vec{F}\rag}\nonumber\\
&=&\sum_{k^j\in\Z}\int\rmd{t_i}\,\mu(\ell,t)\, e^{-\frac{\delta}{4}\sum_{i=1}^M{t}_i{t}_i+\sum_i t_i\lag \hat{e}^i(\ell),\,\vec{F}+2\pi i \sum_j k^j \hat{e}_j^*(\ell)\rag}\label{poisson}
\ee
where $\hat{e}_j^*(\ell)$ is the lattice vector of the lattice $\Fl^*_{NR}(\ell)$ dual to $\Fl_{NR}(\ell)$, satisfying $\langle  \hat{e}^i(\ell), \hat{e}_j^*(\ell)\rangle=\delta^i_j$.

We make a short-hand notation by 
\be
\lag \hat{e}^i,\vec{F}+2\pi i \sum_j k^j \hat{e}_j^*\rag\equiv \Phi_{(k)}^i\equiv i\psi_{(k)}^i e^{i\phi_{(k)}^i},
\ee
where $\psi_{(k)}^i\in\R,\ \phi_{(k)}^i\in[0,2\pi)$. The quantities $\Phi_{(k)}^i,\psi_{(k)}^i,\phi_{(k)}^i$ depend on $\ell,g_{ve},z_{vf}$. We perform the Gaussian integral of ${t}$: 
\be
&&\int\rmd{t}_i\,\mu(\ell,t)\, e^{-\frac{\delta}{4}\sum_{i=1}^M{t}_i{t}_i+\sum_{i=1}^M t_i\Phi^i_{(k)}}\nonumber\\
&=&2^{N_f}\lt({\frac{4\pi}{\delta}}\rt)^{\frac{M}{2}}\prod_f\lt({J}_f(\ell)+\sum_{i=1}^M (\hat{e}^i)_f\frac{\partial}{\partial\Phi^i_{(k)}}\rt)e^{\sum_{i=1}^M\frac{1}{\delta}\Phi^i_{(k)}\Phi^i_{(k)}}\nonumber\\
&=&2^{N_f}\lt({\frac{4\pi}{\delta}}\rt)^{\frac{M}{2}}\prod_f\lt({J}_f(\ell)+\sum_{i=1}^M \frac{2}{\delta}\Phi_{(k)}^i(\hat{e}^i)_f\rt)e^{\sum_{i=1}^M\frac{1}{\delta}\Phi^i_{(k)}\Phi^i_{(k)}}\nonumber\\
&\equiv& D_\delta^{(k)}(\ell,g_{ve},z_{vf})
\ee
The spinfoam amplitude now reads,
\be
Z_{\sn,\delta}
=\sum_{\vec{J}(\ell)}\int \rmd g_{ve}\rmd z_{vf}\, e^{\lag \vec{J}(\ell),\vec{F}\rag}\sum_{\{k^j\}\in\Z^M}D_\delta^{(k)}(\ell,g_{ve},z_{vf}).\label{ZN1}
\ee
The regulator $\delta$ defines a deformation from the original spinfoam amplitude $Z_{\sn}$.


As it becomes clear in the next section, when $F_f$ is restricted to be purely imaginary, $\Phi^i_{(k)}=i\psi_{(k)}^i\in i\R$. Then $D^{(k)}_\delta$ reduces to
\be
D^{(k)}_\delta(\ell, g_{ve},z_{vf})
&=&\lt(\frac{4\pi}{\delta}\rt)^{\frac{M}{2}}e^{-\frac{1}{\delta}\sum_{i=1}^M\psi^i_{(k)}\psi^i_{(k)}}\nonumber\\
&&\!\!\!\!\!\!\!\!\!\!2^{N_f}\prod_f\lt({J}_f(\ell)+\frac{2i}{\delta}\sum_{i=1}^M \psi^i_{(k)}(\hat{e}^i)_f\rt).
\ee
As $\delta\to0$, $D^{(k)}_\delta$ contains a gaussian peaked at $\psi^i_{(k)}=0$ with width $\sqrt{\delta}$. Its center $\psi^i_{(k)}=0$ means 
\be
\lag \hat{e}^i,\vec{F}+2\pi i \sum_j k^j \hat{e}_j^*\rag=\lag \hat{e}^i,\vec{F}\rag+2\pi i\, k^i =0
\ee
The sum over $\{k^j\}\in\Z^M$ in Eq.\Ref{ZN1} reflects that $Z_\sn$ is periodic in ${F}_f\to F_f +4\pi i$. The above peakedness of $D^{(k)}_\delta$ and the sum over $\{k^j\}$ is a consequence of the periodicity.



\section{Regge Equation and Small Deficit Angle}\label{RESDA}

The amplitude $Z_{\sn,\delta}$ depends on 2 independent scales $(\l,\delta)$, where (1) $\l$ is the mean value of $\tilde{J}_f\equiv\l j_f$ in $\sn_{Regge}\subset\sn$, and (2) $\delta$ is the regulator in $D^{(k)}_\delta$ for regulating the transverse $\vec{t}$-sum of non-Regge-like spins. Here $\l\gg 1$ since we are interested in large-$J$ regime, while $\delta\ll1$ since the regulator should be turned off in the end. However we may let 2 scaling limits $\l\to\infty$ and $\delta\to 0$ compete, to find an physically interesting regime. 

$\l$ relates to the length scale where the semiclassical expansion of spinfoam amplitude is defined, since the typical lattice spacing is $\ell\sim(\l\g\ell_P^2)^{1/2}$ for geometries in $\sn$. It turns out the other parameter $\delta$ relates to the continuum limit in refining the lattice. $\delta$ provides a bound to ensure the lattice spacing $\ell$ is always much smaller than the typical curvature radius $\rho$ in all geometries emergent from spinfoam amplitude. It guarantees the simplicial geometries to approach the continuum in the lattice refinement.

It turns out that an interesting way of arranging limits is to first take $\l\to\infty$ then $\delta\to 0$. In other words, the interesting regime is that $\l\gg 1/\delta\gg1$

When we first take the asymptotical limit $\l\to\infty$, $D_\delta$ doesn't oscillate or suppress, thus doesn't affect critical equations from $\langle \vec{J}(\ell),\vec{F}\rangle$. When $\vec{J}(\ell)=\l \vec{j}(\ell)$ represents Regge-like spins, there alway exist solutions to critical equations 
\be
\mathrm{Re}\vec{F}=\partial_g\langle \vec{j}(\ell),\vec{F}\rangle=\partial_z\langle \vec{j}(\ell),\vec{F}\rangle=0, 
\ee
Solutions $(j_f(\ell),g_{ve}(\ell),z_{vf}(\ell))$ correspond to nondegenerate Regge geometries on $\ck$, parametrized by the edge-lengths ${\ell}$ which relates $\vec{J}$ by Eq.\Ref{area-length}. There may not be a unique set of $\ell$ corresponding to a given Regge-like $\vec{J}$. If it happens, critical solutions contains different Regge geometries with different sets of edge lengths.

Note that when $\vec{J}(\ell)$ is a representative away from $\sn_{Regge}$ with $O(1)$ distance, $(j_f(\ell),g_{ve}(\ell),z_{vf}(\ell))$ are approximate solutions to the critical equations with $O(1/\l)$ errors.

Given a set of edge-lengths $\vec{\ell}$ of a nondegenerate Regge geometry, in principle it corresponds to $2^{N_\sig}$ critical solutions ($N_\sig$ is the number of 4-simplices), which has indefinite local 4d orientations at each 4-simplex $\sig$ \cite{semiclassical,HZ}\footnote{This result is valid for the Lorentzian spinfoam amplitude. The Euclidean amplitude gives $4^{N_\sig}$ critical solutions instead of $2^{N_\sig}$. There are 4 solutions $(g_{ve},g'_{ve})$, $(g'_{ve},g_{ve})$, $(g_{ve},g_{ve})$, $(g'_{ve},g'_{ve})$ in each 4-simplex. But different critical solutions are still understood as belonging different well-separated sectors, as in the Lorentzian case. Again we only consider the sector of $g^+_{ve}\neq g^-_{ve})$ with a global orientation. }. Within $2^{N_\sig}$ solutions, there are 2 solution corresponding to 2 different global orientations. Here we only concern about the sector of critical solutions corresponding to globally oriented Regge geometries. Small perturbations doesn't flip the 4-simplex orientation, thus doesn't relate solutions from different sectors\footnote{The 4-simplex orientation only takes discrete values $\pm1$ \cite{HZ}. Small deformations among critical solutions doesn't affect the value of orientation.}. We are going to determine whether the critical solutions in the sector give dominant contribution to the spinfoam amplitude in the regime $\l\gg 1/\delta\gg1$. It turns out that a subset of critical solutions indeed give the leading contribution to the amplitude. As is shown in the following, among critical solutions in this sector, the dominant contribution of spinfoam amplitude comes from the critical solutions whose corresponding Regge geometries are of small deficit angle $\eps_f\ll 1$ and satisfying the Regge equation.


At critical solutions with global orientation, the asymptotical limit $\l\to\infty$ gives
\footnote{Note that at each $\{\ell\}$ in $\sum_\ell$ in Eq.\Ref{Zregge}, the critical solutions beyond the above sector may contribution some exponentials in addition to $e^{{i}S_{Regge}[\ell]/{\ell_P^2}+\cdots}$. If we denote by $\sig$ all possible assignment of orientations to simplices ($\sig$ also includes the solutions with $g^+_{ve}=g^-_{ve}$ in Euclidean amplitude), the asymptotical behavior Eq.\Ref{Zregge} of $Z_{\sn,\delta}$ may be more properly written as
\be
\sum_\sig\sum_{\ell} e^{\frac{i}{\ell_P^2}S_\sig[\ell]+\cdots}\sum_{\{k^j\}\in\Z^M}D^{(k)}_{\delta,\sig}(\ell, g_{ve}(\ell),z_{vf}(\ell))\label{sigmas}
\ee 
Each ${i}S_\sig[\ell]/{\ell_P^2}$ is the spinfoam action evaluated at the critical solution with orientations $\sig$ in simplices. Eq.\Ref{Zregge} corresponds to the term where $\sig$ endows $\ck$ a global orientation. The leading contributions to $Z_{\sn,\delta}$ in Eq.\Ref{sigmas} have been organized into disjoint sectors associated to different $\sig$. Each sector $\sig$ has its own partition function $\sum_{\ell} e^{iS_\sig/{\ell_P^2}+\cdots}\sum_{k^j}D^{(k)}_{\delta,\sig}$. Small perturbations don't relate critical solutions from different sectors. In other words, those critical solutions without global orientation only give non-perturbative corrections to Eq.\Ref{Zregge}. In this paper, we focus on the sector in Eq.\Ref{sigmas} with a global orientation, and study the geometries making leading contributions to the amplitude.}, 
\be
Z_{\sn,\delta}\sim\sum_{\ell} e^{\frac{i}{\ell_P^2}S_{Regge}[\ell]+\cdots}\sum_{\{k^j\}\in\Z^M}D^{(k)}_\delta(\ell, g_{ve}(\ell),z_{vf}(\ell)).\label{Zregge}
\ee 
We have replace $\sum_{\vec{J}(\ell)}$ by $\sum_{\ell}$, since critical solutions contains all possible $\ell$ relating to $\vec{J}$. $S_{Regge}$ is the Regge action
\be
S_{Regge}[\ell]=\sum_{f} {\mathbf{a}}_f {\eps}_f+\sum_{f\subset\partial\ck} {\mathbf{a}}_f {\Theta}_f,\quad {\mathbf{a}}_f=\g \tilde{J}_f(\ell) \ell_P^2
\ee
where $\tilde{J}_f(\ell)\in \sn_{Regge}$ has been represented by its nearest neighbor $J_f(\ell)$. Here $\cdots$ stands for the subleading corrections in large-$J$. 

In the above asymptotical behavior, $S_{Regge}[\ell]$ is obtained by evaluation of $\langle \vec{\tilde{J}}(\ell),\vec{F}\rangle$ at the critical solution corresponding to the Regge geometry $\{\ell\}$. $F_f$ evaluated at the critical solution gives $i\g\eps_f$ at each internal $f$ and gives $i\g\Theta_f$ at each boundary $f$, where $\eps_f$ and $\Theta_f$ are the bulk deficit angle and boundary dihedral angle in the Regge geometry. See \cite{HZ} for the detailed derivation.

At the leading order, $D^{(k)}_\delta$ takes value at the critical solution $g_{ve}(\ell),z_{vf}(\ell)$. At each critical point, $\mathrm{Re}\vec{F}=0$, and $F_f=i\g\eps_f$ for each internal $f$. Thus $\Phi^i_{(k)}\in i\R$, and  
\be
D_\delta^{(k)}(\ell, g_{ve}(\ell),z_{vf}(\ell))&=&\lt(\frac{4\pi}{\delta}\rt)^{\frac{M}{2}}e^{-\frac{1}{\delta}\sum_{i=1}^M\psi^i_{(k)}(\ell)\,\psi^i_{(k)}(\ell)}\nonumber\\
&&\!\!\!\!\!\!\!\!\!\!\!\!\!\!\!\!\!\!\!\!\!\!\!\!\!\!\!\!\!\!\!\!\!\!\!2^{2N_f}\prod_f\lt({J}_f(\ell)+\frac{2i}{\delta}\sum_{i=1}^M \psi^i_{(k)}(\ell)\,(\hat{e}^i)_f(\ell)\rt),
\ee
where 
\be
\psi^i_{(k)}(\ell)=\g\lag\hat{e}^i,\vec{\eps}\rag+2\pi k^i.
\ee
Because of the gaussian $e^{-\frac{1}{\delta}\sum_{i=1}^M\psi^i_{(k)}\psi^i_{(k)}}$ with small $\delta$, each $D^{(k)}_\delta$ is essentially supported within a small neighborhood of size $\sqrt{\delta}$ at $\psi^i_{(k)}=0$. As $\delta\ll1$, each $D_\delta$ effectively suppresses the contributions from configurations with large $\psi^i_{(k)}$, and picks out the configurations with small $\psi^i_{(k)}$.

As the large-$J$ limit $\l\to\infty$ gives $\ell_P^2\ll \Ar_f$, from the variational principle (see Appendix \ref{EMF}), the leading contribution of Eq.\Ref{Zregge} is given by the $\{\ell\}$ configurations satisfying Regge equation
\be
\sum_f\frac{\partial\Ar_f }{\partial\ell}\eps_f=0,\quad \text{or}\quad \g\lag\frac{\partial \vec{J} }{\partial\ell},\,\vec{\eps} \rag=0.\label{ER}
\ee
Each solution of Regge equation gives the leading order contribution to $Z_{\sn,\delta}$, which is proportional to
\be
e^{\frac{i}{\ell_P^2}\sum_{f\subset\partial\ck} {\mathbf{a}}_f {\Theta}_f}\sum_{\{k^j\}\in\Z^M} e^{-\frac{1}{\delta}\sum_{i=1}^M\psi^i_{(k)}(\ell)\,\psi^i_{(k)}(\ell)}(\cdots).\label{contribution}
\ee
Note that the bulk terms in $S_{Regge}[\ell]$ vanishes at each solution of Regge equation. Now we take $\delta\ll1$, the Gaussian $e^{-\frac{1}{\delta}\sum_{i=1}^M\psi^i_{(k)}\psi^i_{(k)}}$ suppresses the amplitude contributed by the solutions $\{\ell\}$, which have relatively large $\psi^i_{(k)}(\ell)=\g\lag\hat{e}^i,\vec{\eps}\rag+2\pi k^i$, i.e. the essential contribution of the spinfoam amplitude $Z^{(\vec{k}=0)}_{\sn,\delta}$ comes from the solutions $\{\ell\}$ satisfying
\be
\lt|\g\lag\hat{e}^i,\vec{\eps}\rag\rt|\leq\delta^{1/2}\ll1\ \text{mod}\ 2\pi k^i.\label{suppress}
\ee

Let's temporarily ignore the terms with $k^j\neq 0$ in Eq.\Ref{contribution}. ${\partial \vec{J} }/{\partial\ell}$ are tangent vectors on the submanifold $\sm_{Regge}$ of Regge-like spins. Thus ${\partial \vec{J} }/{\partial\ell}$ and $\hat{e}^i$ form a complete basis in $\sn$. The Regge equation Eq.\Ref{ER} and the requirement Eq.\Ref{suppress} at $k^j=0$ combine and give that all deficit angles have to be small
\be
|\g\eps_f|\leq\delta^{1/2}\ll1.\label{smalleps}
\ee
Namely, given a solution $\{\ell\}$ to Regge equation, all its deficit angles $\eps_f$ have to be small in order to provide a non-suppressed contribution to the spinfoam amplitude at $k^j=0$. The Barbero-Immirzi parameter $\g$ is a fixed $O(1)$ parameter in our discussion. If $\g$ was not fixed and sent to zero combining the semiclassical limit, Eq.\Ref{smalleps} would allow large deficit angle in the semiclassical Regge geometries, which reproduced the result in \cite{claudio,lowE2}. 

When the simplicial triangulation is refined, given a Regge geometry $\{\ell\}$ which approximate a smooth geometry\footnote{If we embed the Regge geometry in $\R^N$, $N>4$, the corresponding smooth geometry is an smooth enveloping surface $\mathscr{S}$ of the Regge geometry, where all vertices (end points of $\ell$'s) in the Regge geometry are located in $\mathscr{S}$. $\mathscr{S}$ is required to satisfy $\rho\gg\ell$ everywhere. Once a $\mathscr{S}$ is chosen, the Regge geometry is a piecewise linear approximation to $\mathscr{S}$ satisfying $|\ell/\ell_s -1|\simeq O({\ell^2}/{\rho^2})$ where $\ell_s$ is the geodesic length connecting the end points of $\ell$ \cite{FFLR}.}, the deficit angle relates to the typical lattice spacing $\ell$ of the Regge geometry and the typical curvature radius $\rho$ of the smooth geometry by \cite{FFLR}\footnote{Given a small 2-face $f$ embedded in a smooth geometry, the loop holonomy of spin connection along $\partial f$ gives $e^{\eps\hat{X}}$, where $\hat{X}$ is the bivector tangent to $f$. As $f$ is small, the holonomy gives $1+\int_f F\simeq 1+\eps\hat{X}$, which implies $\eps\simeq \ell^2/\rho^2$ since $F$ is the curvature 2-form of the spin connection. Typical spacings of $\ck$ and $\ck^*$ are of similar scales. }
\be
\eps\sim \frac{\ell^2}{\rho^2}\lt[1+ O\lt(\frac{\ell^2}{\rho^2}\rt)\rt].\label{ellrho}
\ee
The Regge geometry has to satisfy $\ell^2\ll\rho^2$ in order to approximate the smooth geometry, since the ratio between $\ell$ and a geodesic length $\ell_s$ of the smooth geometry is $\ell/\ell_s=1+O({\ell^2}/{\rho^2})$. Note that the smooth limit of Regge geometry also requires the fatness of simplices is bounded away from zero, to avoid any degenerate simplex. See e.g. \cite{BARRETT1994107,cheeger1984,BarrettLinRegge} for details.   

When the lattice is sufficiently refined, and when $\delta$ is sent to be small, Regge geometries sufficiently approximate smooth geometries all satisfy Eq.\Ref{smalleps} and survive as dominant contribution to $Z_{\sn,\delta}$ at $k^j=0$. Regge geometries suppressed by $D_\delta$ are the ones which fail to approximate any smooth geometry. The regulator $\delta$ behaves similarly as the bound of error in the piecewise linear approximation of smooth metric 
\be
\lt|\ell/\ell_s-1\rt|\simeq O({\ell^2}/{\rho^2})\leq \delta^{1/2}.\label{smoothness}
\ee

The leading contribution to the semiclassical spinfoam amplitude must satisfy both Regge equation \Ref{ER} and Eq.\Ref{smalleps}. Therefore the solutions of Regge equation which approximate smooth geometries all give dominant contributions to the spinfoam amplitude.


The terms with $k^j\neq 0$ add discrete ambiguities to the constraint Eq.\Ref{smalleps}. However different $k^j$ correspond to disjoint sectors of discrete geometries satisfying Eq.\Ref{suppress}. Geometries in sectors of $k^j\neq0$ don't approximate any smooth geometry. Small perturbations cannot relate two geometries satisfying Eq.\Ref{suppress} with different $k^j$.

The geometries in sectors with $k^j\neq 0$ may have non-suppressed contributions to the semiclassical spinfoam amplitude (as has been pointed out in \cite{lowE,LowE1}). However the sectors is sensitive to the choice of the neighborhood $\sn_{Regge}$ in defining $Z_{\sn,\delta}$. For example, we assume the neighborhood $\sn_{Regge}$ which contains the physical Regge geometries only with relatively small deficit angles, i.e. $\g\langle \hat{e}^i,\vec{\eps}\rangle$ is not close to any $2\pi k^i$ with $k^i\neq 0$. Then the terms with $k^j\neq 0$ in Eq.\Ref{contribution} only have negligible contribution to $Z_{\sn,\delta}$. The dominant contribution to $Z_{\sn,\delta}$ comes from the geometries with small deficit angles. $k^j= 0$ sector is physically most relevant because it is the only sector containing discrete geometries approaching the continuum as the simplicial lattice being refined.

It is mentioned in Section \ref{1} that critical points in the spinfoam action contain time non-oriented geometries \cite{hanPI}, which gives $F_g=i(\g\eps_f\pm\pi)$. Within this type of critical points, the equation of motion Eq.\Ref{ER}, the contraint Eq.\Ref{suppress} or Eq.\Ref{smalleps}, are modified by the replacement $\g\eps_f\to\g\eps_f\pm\pi$. The constraint then leads to that $\g\eps_f$ is close to $\pm\pi$. These critical points form 2 disjoint sectors away from the ones discussed above. Geometries in this sector doesn't approximation any smooth geometry, and can be treated in the same way as the $k^j\neq 0$ sectors. Some discussion of the Euclidean amplitude is given in Appendix \ref{Eucl}.

\section{Einstein-Regge Regime}\label{4ERR}

We refer to the regime of spinfoam model, where the Regge equation emerges together with the constraint $\g\eps_f\leq\delta^{1/2}$, as the \emph{Einstein-Regge (ER) regime}. The ER regime is defined by considering the deformed spinfoam amplitude $Z_{\sn,\delta}(\ck)$, and imposing the following requirements on the parameters $\ck,\sn,\delta$:

\begin{itemize}

\item The neighborhood $\sn$ contains a submanifold $\sn_{Regge}\subset \sn$. All $\tilde{J}_f(\ell)$ in $\sn_{Regge}$ are large $\tilde{J}_f(\ell)\gg1$. The mean value of $\tilde{J}_f(\ell)$ in $\sn_{Regge}$ is denoted by $\l$. Parameters $\l$ and $\delta$ satisfy $\l\gg\delta^{-1}\gg1$

\item The neighborhood $\sn$ of the spinfoam spin-sum has to be compatible with the triangulation $\ck$. Namely, Regge geometries $\{\ell\}$ in the neighborhood $\sn_{Regge}\subset \sn$ all have relatively small deficit angles $\eps_f$ (e.g. requiring $\g{\eps}_f<\pi$). $\sn_{Regge}$ should contain Regge geometries that approximate smooth geometries.


\end{itemize}

In the ER regime specified by the above requirements, the spinfoam amplitude obtains dominant contributions from Regge geometries in $\sn$, which satisfy both the Regge equation \Ref{ER} and the bound $\eps_f\leq\g^{-1}\delta^{1/2}$. These Regge geometries contain the ones approximates smooth geometries by Eq.\Ref{smoothness}. They satisfy the following (approximate) bound by Eq.\Ref{ellrho}
\be
\rho^2\geq\frac{\g \l\ell_P^2}{\sqrt{\delta}}\gg\ell^2\gg\ell_P^2
\ee 
The inequality $\ell_P^2\ll\ell^2\ll\rho^2$, satisfied by the dominant configurations, is the condition that the discrete geometry is semiclassical ($\ell^2\gg\ell_P^2$), as well as approaching the continuum limit ($\ell^2\ll\rho^2$) \cite{lowE,Han:2016fgh,HanHung}. 

It is anticipated that geometries both satisfying Regge equation and approximating the continuum should approximate the smooth solution to the \emph{continuum} Einstein equation. We will come back to this point in the next section.

Note that in this work, we limit ourselves to understand the dominance in spinfoam amplitude from classical geometries with a global orientation. As it has been mentioned in the last section, geometries without global orientation live in other well-separated sectors. They may provide non-pertrubative corrections to the contribution studied above, although they don't affect the perturbative expansion at any classical geometry.

\section{Semiclassical Continuum Limit}\label{SCL}

So far the discussion is based on a fixed triangulation. We may change our viewpoint and consider a sequence of triangulations $\ck_n$, where each $\ck_{n+1}$ is a refinement of $\ck_n$. The vertices of all $\ck_n$'s is a dense set in the manifold where the triangulations are embedded.
The sequence of $\ck_n$ defines a sequence of spinfoam amplitudes $Z_{\sn,\delta}(\ck_n)$. The smooth geometry can be understood as the limit of a sequence of discrete geometries $\{\ell_n\}$ on the sequence of triangulations $\ck_n$, where the discrete geometries approach toward $\ell^2/\rho^2\to 0$. When each of discrete geometries $\{\ell_n\}$ in the sequence satisfies the Regge equation on $\ck_n$, it gives the non-suppressed contribution to the spinfoam amplitude $Z_{\sn,\delta}$ on $\ck_n$. 

Let's come into more detailed behavior of geometries $\{\ell_n\}$ and amplitudes $Z_{\sn,\delta}$ on the sequence of triangulations $\ck_n$. Generically on a more refined triangulation, the large system size requires a larger $\l$ to obtain the semiclassical behavior as the leading order in the spinfoam amplitude. Indeed in the $1/\l$ quantum correction of the amplitude, the coefficient of $1/\l^s$ is a sum over all $g_{ve},z_{vf}$ degrees of freedom on the triangulation (see e.g. \cite{hanPI}).
\be
i^{-s}\sum_{l-m=s}\sum_{2l\geq 3m}\frac{2^{-l}}{l!m!}\lt[\sum_{a,b}H^{-1}_{ab}(x_0)\frac{\partial^2}{\partial x_a\partial x_b}\rt]^lg_{x_0}(x_0)^m\label{Lsu}
\ee
where $x_0$ is a critical point, $H(x)=S''(x)$ denotes the Hessian matrix, and $g_{x_0}(x)$ is given by 
\be
g_{x_0}(x)=S(x)-S(x_0)-\frac{1}{2}H^{ab}(x_0)(x-x_0)_a(x-x_0)_b. 
\ee
Here $a,b$ label all degrees of freedom on the triangulation. A refined triangulation carries a larger number of degrees of freedom, thus generically produce a larger coefficient. It requires a smaller $1/\l$ to suppressed the quantum correction and let the semiclassical behavior stand out. Therefore the discrete geometry $\{\ell_n\}$ on $\ck_n$ has larger and larger $\l$ as $\ck_n$ becomes more and more refined. Even if it happens that the above generic behavior is violated in certain situation, i.e. the coefficient of $1/\l$ doesn't increase in refining the lattice, tuning $\l$ larger still suppresses the quantum correction. So $\l$ can in general set to be monotonically increasing in refining the lattice.  

Naively it might sound unexpected to have $\l$ larger in the refinement since the triangle area $\ell^2\sim\Ar=\g \l\ell_P^2$. However the continuum limit is controlled by the ratio $\ell^2/\rho^2$. The ratio becomes smaller when the curvature radius $\rho$ in Planck unit increases in a faster rate than $\l$, or equivalent, when we zoom out to larger length unit such that the value of $\ell_P$ decrease in a faster rate. Zooming out to larger length unit is required by the semiclassical limit.

Formally we associate each triangulation $\ck_n$ a mass scale $\mu_n$ whose inverse $\mu_n^{-1}$ is a length unit. $n$ becoming larger is the refinement of $\ck_n$, while $\mu_n$ becomes smaller. The length unit $\mu_n^{-1}$ increase as refining the triangulation. Given the 1-to-1 correspondence between $\ck_n$ and $\mu_n$, we may simply label the triangulation and discrete geometry as $\ck_\mu$ and $\{\ell_\mu\}$ by its associated scale $\mu$. $\ck_\mu$ is refined as $\mu$ going to infrared (IR). On each $\ck_\mu$, the discrete geometry gives the triangle area $\Ar(\mu)$ 
\be
\Ar(\mu) = \g \l(\mu)\ell_P^2= a(\mu)\mu^{-2}.\label{amu5}
\ee
Here the running of $\ell_P$ is not considered since we are in the semiclassical limit. $\l(\mu)$ increases monotonically in the refinement $\mu\to 0$ as discussed above. However we can assign the scale $\mu$ to $\ck_\mu$ such that $a(\mu)\to 0$ as $\mu\to 0$ \footnote{Considering the gap $\Delta J_f=\half$, $\Delta a_f(\mu)=\g\Delta J_f(\mu)\mu^2\ell_P^2=\frac{1}{2}\g\mu^2\ell_P^2 \to 0$ as $\mu\to0$}. Using the dimensionless length $a(\mu)$, we can define the convergence of the sequence of geometries $\{\ell_\mu\}$ ( where $\ell_\mu\sim {a(\mu)}^{1/2}\mu^{-1}$ for each geometry) on $\ck_\mu$ converge to a smooth geometry by requiring $\lim_{\mu\to0}a(\mu)=0$ and the fatness bounded away from zero. The target smooth geometry has the dimensionless curvature radius denoted by $L$, which is the curvature evaluated at the IR unit $\mu\to 0$, i.e. the dimensionful curvature radius is
\be
\rho(\mu)= L\mu^{-1}.
\ee
The sequence of discrete geometries $\{\ell_\mu\}$ approaches the smooth geometry because 
\be
\frac{\Ar(\mu)}{\rho(\mu)^2}=\frac{a(\mu)}{L^2}\to 0,\quad \text{as}\quad \mu\to 0.
\ee
Note that since $\mu$ is of mass dimension, $\mu\to 0$ may be understood more appropriately as $\mu\ell_P\to0$.

The dependence of $\l$ on $\mu$ shows that the semiclassical limit is taken at the same time as the lattice refinement limit. Possible assignments of scales $\mu$ to triangulations $\ck_\mu$ are classified in Section \ref{running}.

As an illustration of the above idea, let's consider a smooth sphere with a unit curvature radius $L=1$. It is standard to define discrete geometries on a sequence of refined triangulations of the sphere, which approaches the smooth sphere in the continuum limit. We assign a mass scale $\mu$ to label the triangulation $\ck_\mu$ such that the refinement relates to $\mu\to0$. On each $\ck_\mu$, edge lengths are $\sqrt{a(\mu)}$ satisfying $\lim_{\mu\to 0}a(\mu)=0$. $\sqrt{a(\mu)}$ are understood as edge lengths in the unit $\mu^{-1}$. The scale $\mu$ should be chosen such that $a(\mu)\mu^{-2}/\ell_P^2\to\infty$ as $\mu\to 0$, in order to have $\l(\mu)$ increasing in the refinement. Geometries in the sequence now associate with different scales $\mu$. The smooth sphere lives at the IR limit whose curvature radius $L=1$ is measured at the IR unit $\mu^{-1}\to\infty$.

Let's turn to the semiclassical behavior of $Z_{\sn,\delta}$ on the sequence of $\ck_\mu$. Here $\sn$ depends on $\mu$ since $\l$ does. We take $\sn(\mu)$'s satisfy the requirement of ER regime. Then $\sn(\mu)$'s contain sequences of Regge geometries which converge to smooth geometries, since $a(\mu)\to 0$ as $\mu\to0$. Moreover since $\l(\mu)$ increases as $\mu\to0$. The existence of ER regime $\l(\mu)\gg \delta^{-1}\gg1$ can be achieved by smaller $\delta$, if we make $\delta=\delta(\mu)$ run with the scale. Namely, we can make $\delta(\mu)\to 0$ as $\mu\to 0$, while $\l(\mu)\gg \delta(\mu)^{-1}\gg1$ is satisfied. For sequences of discrete geometries $\{a(\mu)^{1/2}\}$ converging to smooth geometries at IR, they give dominant contributions to $Z_{\sn(\mu),\delta(\mu)}$ at each $\mu$, if they satisfy Regge equation on each $\ck_\mu$ and
\be
\g\frac{a(\mu)}{L^2}\leq \delta(\mu)^{\half}.
\ee
We may choose decreasing rates of $\delta(\mu)^{1/2}$ and $a(\mu)$ to be the same, to keep all converging geometries contributing dominantly. $\delta(\mu)\to 0$ as $\mu\to 0$ means that the regulator $\delta$ is removed in the continuum limit, where $Z_{\sn,\delta}$ goes back to its original definition Eq.\Ref{Z}.

Spinfoam amplitudes give sequences of Regge geometries converging to smooth geometries, where each geometry satisfies the Regge equation on its lattice. It is thus expected that each smooth geometry as the limit is a solution of continuum Einstein equation. However due to complexities of both Regge equation and Einstein equation, a general mathematical proof is unfortunately not available in the literature as far as we know. However there have been extensive studies on the continuum limit of Regge calculus, which gives many analytic and numeric examples (see \cite{Williams:1991cd,Gentle:2002ux} for summaries). In all the examples, solutions of Regge equation always converge to smooth solutions to Einstein equation. Among the examples, there have been constructions of solutions of linearized Regge equations in Euclidean signature, which converge to solutions to linearized Einstein equation \cite{0264-9381-5-12-007,0264-9381-5-9-004,Christiansen2011}. In the nonlinear regime, there have been numerical simulations of time evolutions in Regge calculus in Lorentzian signature, as a tool of numerical relativity. Nontrivial results include e.g. Kasner universe, Brill waves, binary black holes, FLRW universe \cite{Gentle:2012tc,Gentle:1997df,Gentle:2002ux,Gentle:1998qg,Liu:2015gpa}. A key observation in the convergence results is that the deviation of Regge calculus from general relativity is the non-commutativity of rotations in the discrete theory, while the error from the non-commutativity is of higher order in edge lengths \cite{Gentle:2008fy}. There is also the convergence result by certain average of Regge equations \cite{Miller:1995gz}. The existing results all demonstrate that Regge calculus is a consistent second order accurate discretization of general relativity.

Given any sequence of solutions to the Regge equation which converges to a solution to the continuum Einstein equation, our analysis shows that each solution gives the dominant contribution to the spinfoam amplitude on $\ck_\mu$ in the semiclassical limit. The smooth solution to Einstein equation is the limit of a sequence of dominant configurations from spinfoam amplitude. 

As an example, Euclidean spinfoam amplitudes on $\ck_\mu$ can give a sequence of solutions to linearized Regge equations, which coincide with the ones constructed in \cite{0264-9381-5-12-007}. Edge lengths used there should be identified with $\sqrt{a(\mu)}$ (more precisely, relates to $a(\mu)$ by Eq.\Ref{area-length}). The sequence of geometries provide dominant contribution to spinfoam amplitudes, and converge in the IR limit $\mu\to0$ to smooth gravitational waves satisfying linearized Einstein equation.

There is another way to obtain the continuum Einstein equation from the convergence of Regge actions. Let's come back to Eq.\Ref{Zregge} and consider the sequence $Z_{\sn(\mu),\delta(\mu)}(\ck_\mu)$. For each sequence of Regge geometries converge to the smooth geometry as $\mu\to 0$, Regge actions converge to the Einstein-Hilbert action on the continuum, when Regge geometries converge to the smooth geometry (The convergence again requires the fatness of simplices to be bounded from zero in addition to shrinking edge lengths, see \cite{cheeger1984,BARRETT1994107} for details). Translating the known convergence result to our context uses the length unit $\mu^{-1}$. We apply Eq.\Ref{amu5} to the Regge action $\frac{1}{\ell_P^2}\sum_{f}\Ar_f(\mu)\eps_f(\mu)$:
\be
\frac{1}{\mu^2\ell_P^2}\sum_{f}a_f(\mu)\eps_f(\mu)= \frac{1}{\mu^2\ell_P^2}\int\rmd^4x\sqrt{-g}\, R \lt[1+ \epsilon\lt(\mu\rt)\rt]\label{EH}
\ee
Where $\epsilon(\mu)$ satisfies $\lim_{\mu\to0}\epsilon(\mu)=0$. The convergence happens as the edge length $a(\mu)^{1/2}\to 0$ at the IR. Smooth geometries and $\int\rmd^4x\sqrt{-g}\, R $ live at the IR limit $\mu\to0$. $\mu^2\ell_P^2$ is the numerical value of $\ell_P^2$ in the unit $\mu^{-2}$. $\mu^2\ell_P^2$ tends to zero when we zoom out to larger unit. 


Given a Regge geometry $\{\ell\}$ approximating the smooth geometry, there is a smooth enveloping surface $\mathscr{S}$ whose curvature satisfies $\rho\gg\ell$ everywhere, and $|\ell/\ell_s-1|\simeq O(\ell^2/\rho^2)$, as well as the fatness bounded away from zero. Small perturbations at $\{\ell\}$ generically don't break the above properties, so only lead to Regge geometries still approximating smooth geometries. 

Indeed, consider a small perturbation of both the Regge geometry and correspondingly, its smooth enveloping surface $\mathscr{S}'$, i.e. $|l'-l|\leq\delta_1$ and $|l_s'-l_s|\leq\delta_2$ with $0<\delta_{1,2}<l^2< l/2$ ($l$ denotes the edge length in unit $\mu^{-1}$). In \cite{cheeger1984,BARRETT1994107}, the rigorous approximation criterion is $|l-l_s|\leq Cl^2$, which gives $|l'-l'_s|\leq Cl^2+\delta_1+\delta_2<(C+2)l^2\leq C'(l-\delta_1)^2\leq C'l'^2$ for $C'= 4(C+2)> \frac{C+2}{(1-\delta_1/l)^2}$. So the perturbed Regge geometry still satisfy the approximation criterion. 

The vicinity of a Regge geometry approximating the smooth geometry only covers Regge geometries that approximate smooth geometries, so Eq.\Ref{EH} is valid in the vicinity. Considering the vicinity is sufficient for the variational principle. The partition function Eq.\Ref{Zregge} within the vicinity (of each approximated smooth geometry) behaves as
\be
Z_{\sn(\mu),\delta(\mu)}(\ck_\mu)\simeq\int [Dg_{\mu\nu}]\ e^{\frac{i}{\mu^2\ell_P^2}\int\rmd^4x\sqrt{-g}\, R \lt[1+ \epsilon\lt(\mu\rt)\rt]}.\label{ZEH}
\ee
Moreover, Eq.\Ref{ZEH} manifests that the IR limit $\mu\to0$ leads to the stationary phase approximation in Eq.\Ref{ZEH}, whose variational principle gives the continuum vacuum Einstein equation $R_{\mu\nu}=0$.

The above argument shows that the spinfoam amplitude reduces to a partition function of Einstein-Hilbert action in the semiclassical continuum limit.


We remark that in the above analysis, the regulator $\delta$ plays an interesting role by opening a window to allow small nonvanishing deficit angles $\eps_f$ for Regge geometries approximating the continuum. Given a sequence of Regge geometries approaching toward a smooth geometry with nontrivial curvature, the small window of $\eps_f$ allows each Regge geometry in the sequence to have dominant contribution in their corresponding (regularized) spinfoam amplitudes $Z_{\sn,\delta}$.

The above result is achieved by taking an appropriate limit combining $\l\to\infty$ and $\delta\to0$ respect to the requirement $\l\gg\delta^{-1}\gg1$ of ER regime. However if the requirement was violated by sending $\delta\to0$ before $\l\to\infty$, we would lose the window of nonvanishing curvature for each Regge geometry in the sequence. Then there would be no smooth curved geometry as the limit from spinfoam amplitudes. This behavior was the flatness observed in \cite{frankflat,flatness}.

\section{Running Scale}\label{running}

In this section we classify all possible assignments of scales $\mu$ to triangulations $\ck_\mu$. In the above discussion, there are two requirements relevant to assigning scales $\mu$ to triangulations $\ck_\mu$:

\begin{itemize}

\item $\l(\mu)$ always suppresses the growth of the coefficient in \Ref{Lsu} at arbitrary order $s$. 

\item $\l(\mu)\mu^2\propto a(\mu)$ monotonically decreases as $\mu\to 0$.

\end{itemize}

We denote the coefficient \Ref{Lsu} at the order $\l^{-s}$ by $f_s(\mu)$, exhibiting its dependence on triangulation $\ck_\mu$. It is required that $|f_s(\mu)|/\l(\mu)^s$ shouldn't blow up as $\mu\to0$ for all $s$:
\be
0\leq\frac{\rmd}{\rmd\mu}\lt(\frac{|f_s(\mu)|}{\l(\mu)^s}\rt)=-\frac{s|f_s|}{\l^{s+1}}\frac{\rmd \l}{\rmd\mu}+\frac{1}{\l^{s}}\frac{\rmd|f_s| }{\rmd\mu}
\ee
which gives
\be
\frac{1}{\l}\frac{\rmd \l}{\rmd\mu}\leq\frac{1}{s|f_s|}\frac{\rmd|f_s| }{\rmd\mu}.\label{app1}
\ee

On the other hand, monotonically decreasing $\l(\mu)\mu^2\propto a(\mu)$ as $\mu\to0$ implies
\be
0<\frac{\rmd}{\rmd\mu}\lt(\l(\mu)\mu^2\rt)=\mu^2\frac{\rmd\l}{\rmd\mu}+2\mu \l
\ee
which gives
\be
0>\frac{1}{\l}\frac{\rmd\l}{\rmd\mu}>-\frac{2}{\mu}.\label{applambda}
\ee

Combining Eq.\Ref{app1} gives
\be\label{app2}
\frac{1}{s|f_s|}\frac{\rmd|f_s| }{\rmd\mu}> -\frac{2}{\mu}
\ee

Recall that $\mu$ is assigned to a sequence of triangulations $\ck_{n}\equiv\ck_{\mu_n}\equiv \ck_{\mu}$. The variable $\mu\equiv\mu_n$ is actually discrete. $|f_s(\mu)|$ and $\l(\mu)$ have been assumed to be a differentiable function which continue $|f_s(\mu_n)|$ and $\l(\mu_n)$. 

Integrating Eq.\Ref{app2} 
\be
\int^{\mu_{n-1}}_{\mu_n}\frac{1}{s|f_s|}\frac{\rmd|f_s| }{\rmd\mu}\rmd\mu> -\int^{\mu_{n-1}}_{\mu_n}\frac{2}{\mu}\rmd\mu
\ee
which gives
\be
\frac{\mu_{n-1}}{\mu_n}>\lt|\frac{f_s(\mu_n)}{f_s(\mu_{n-1})}\rt|^{\frac{1}{2s}}.\label{app3}
\ee
Thus the assignment of $\mu$ depends on the behavior of coefficients $f_s(\mu_n)$ for all $s$. All possibilities are classified as follows:

\begin{enumerate}

\item The simplest situation is that all $|f_s(\mu)|$ stops increasing at finite $\mu_s>\mu_*>0$, then Eq.\Ref{app3} doesn't impose any constraint to $\mu$ when $\mu<\mu_*$, since ${\mu_{n-1}}/{\mu_n}$ always greater than 1. It is easy to find $\l(\mu)$ to satisfy Eq.\Ref{applambda}.

\item If there are finitely many $s\geq 1$ whose $|f_s(\mu)|$ increase monotonically as $\mu\to0$, finitely many $\lt|\frac{f_s(\mu_n)}{f_s(\mu_{n-1})}\rt|>1$ impose nontrivial lower bound to ${\mu_{n-1}}/{\mu_n}$. Because the number of increasing $|f_s(\mu)|$ is finite, there is a bounded $B_n$ at each $n$
\be
\lt|\frac{f_s(\mu_n)}{f_s(\mu_{n-1})}\rt|^{\frac{1}{2s}}\leq \max_{s\geq1}\lt|\frac{f_s(\mu_n)}{f_s(\mu_{n-1})}\rt|^{\frac{1}{2s}}\equiv B_n.
\ee
We can choose $\frac{\mu_{n-1}}{\mu_n}> B_n$ at each $n$, so that Eq.\Ref{app3} is satisfied uniformly to all orders $s$.

\item If there are infinitely many $|f_s(\mu)|$ increase monotonically as $\mu\to0$, and if the rate $\lt|\frac{f_s(\mu_n)}{f_s(\mu_{n-1})}\rt|\leq A_n e^{C_n s}$ (for certain constants $A_n,C_n>0$) bounded by exponentially growing when going to higher orders $s$. Then there is a upper bound $B_n$ at each $n$ ($A_n^{\frac{1}{2s}}$ is bounded in $s\geq 1$)
\be
\lt|\frac{f_s(\mu_n)}{f_s(\mu_{n-1})}\rt|^{\frac{1}{2s}}\leq A_n^{\frac{1}{2s}} e^{C_n/2}\leq  B_n.
\ee
We can again choose $\frac{\mu_{n-1}}{\mu_n}> B_n$ at each $n$, so that Eq.\Ref{app3} is satisfied uniformly to all orders $s$.

\item If $\lt|\frac{f_s(\mu_n)}{f_s(\mu_{n-1})}\rt|^{\frac{1}{2s}}$ is not bounded from above as $s\to\infty$ at any $n$, Eq.\Ref{app3} can only be satisfied at any truncation of the $\l^{-1}$ asymptotic expansion. At any truncation up to $\l^{-s_0}$ order, $\lt|\frac{f_s(\mu_n)}{f_s(\mu_{n-1})}\rt|^{\frac{1}{2s}}$ at each $n$ is bounded from above within finitely many $1\leq s\leq s_0$. The bound changes for different $s_0$. Then the rate $\frac{\mu_{n-1}}{\mu_n}$ has to be justified order by order. 

\end{enumerate}

We conjecture that the 3rd situation should be most relevant. $f_s$ in quantum mechanics and quantum field theories have the following generic behavior as $s\to\infty$ (see e.g. \cite{Lipatov,Spencer1980,david1988})
\be
|f_s|\simeq \eta s! s^\a \b^s\lt(1+\epsilon(s)\rt)^s,\quad \lim_{s\to\infty}\epsilon(s)=0
\ee  
where constants $\eta,\a,\b$ may depend on different theories and different numbers of degrees of freedom. This behavior leads to that
\be
\lt|\frac{f_s(\mu_n)}{f_s(\mu_{n-1})}\rt|^{\frac{1}{2s}}\simeq\lt(\frac{\eta_n}{\eta_{n-1}}\rt)^{\frac{1}{2s}}s^{\frac{1}{2s}\lt(\a_n-\a_{n-1}\rt)}\lt(\frac{\b_n}{\b_{n-1}}\rt)^\half\lt[1+\epsilon(s)\rt]^\half
\ee
is bounded from above for large $s$.

\section{Conclusion and Outlook}

The discussion of this paper explains the emergence of Einstein equation from spinfoam amplitudes in the semiclassical continuum limit. However the spinfoam amplitude seems contain more solutions than the Einstein equation does. The analysis here mainly focus on the sector of critical points in the spinfoam amplitude which corresponds to nondegenerate geometries with a global orientation. Solutions to the Einstein equation emerge within this sector. There exists other well separated sectors where the spinfoam amplitude gives the degenerate geometry and geometries without a global orientation \cite{semiclassical,HZ}. Those geometries may not satisfy the Einstein equation, and their physical meaning remains open (see e.g. \cite{Christodoulou:2012sm} for some discussion). Note that there exists the spinfoam model (the proper vertex) whose asymptotics give a single orientation to each 4-simplex \cite{Engle:2015zqa}. The discuss in this paper is also valid in this model. 

A key step in the discussion is the regularization of the non-Regge-like spin sum in Eq.\Ref{deform1}, which is a deformation of the spinfoam amplitude. We take the point of view that the spinfoam amplitude defined on a triangulation might be an effective theory from a complete LQG theory as the ``continuum limit'' of the spinfoam. As the level of effective theory, the deformation has to be implemented to spinfoam amplitude to reproduce the desired semiclassical limit. As is shown in Section \ref{SCL}, the deformation is turned off in the continuum limit. It suggests that the spinfoam amplitudes, with or without the deformation, should have the same ``continuum limit''. The amplitude with the deformation is one effective description of the complete LQG theory, whose advantage is the correct semiclassical behavior. 

Although the regularization includes a Gaussian damping factor in the non-Regge-like spin sum, it is not allowed to completely remove non-Regge-like spins in the spin sum. Removing all non-Regge-like spins would be an ad hoc modification of the model, which modified the continuum limit. The modification would remove the small-$\eps_f$ constraint \Ref{smalleps0} or \Ref{smalleps} and break the desired behavior of spinfoam amplitude near a classical curvature singularity in \cite{Han:2016fgh} (reviewed briefly at the end of Section \ref{intro}). In our opinion, the existence of non-Regge-like spins and its consequence, the flatness, are nice properties of spinfoam amplitude, when treated properly.

There has been recent progress on the spinfoam amplitude with cosmological constant \cite{HHKR,HHKRshort,3dblockHHKR,hanSUSY,Han:2017geu,Han:2016dnt,curvedMink}. A research undergoing is to apply the present analysis to the formalism with cosmological constant. Another possible future direction is to apply the analysis to the sum over triangulations in group field theory (GFT). The method developed in this work might be helpful to understand the emergence of classical geometries from GFT, and the relation to phase transitions. Our results on the spinfoam amplitude might also be applied to the tensor network approach in the bulk-boundary duality \cite{Qi1,Pastawski:2015qua}, by the relation between random tensor networks and spin-networks \cite{HanHung}. The recent work in \cite{Han:2017uco} applies discrete 3d bulk gravity to random tensor networks, and reproduces correctly the holographic R\'enyi entropy of 2d CFT. The result here may be useful in the generalization to 4 bulk dimensions. 

Finally we mention that there have been earlier studies on the continuum limit in spinfoams e.g. \cite{Delcamp:2016dqo,Dittrich:2016tys,Dittrich:2013voa,Bahr:2012qj,Bahr:2017klw,Bahr:2016hwc,Gambini:2008as}. There are also some recent results on emerging classical spacetimes from GFT e.g. \cite{Oriti:2016ueo,Oriti:2015rwa,Oriti:2016acw}.





\section*{Acknowledgements}

The author thanks Antonia Zipfel and Warner Miller for helpful discussions, and thanks John Barrett for email communications. He also acknowledges support from the US National Science Foundation through grant PHY-1602867, and Start-up Grant at Florida Atlantic University, USA.

\appendix

\section{Spin Sum in Spinfoam Amplitude}\label{free}

In this section, we show that $\sum_{J_f}$ in the spinfoam amplitude Eq.\Ref{Z} can be understood as a free spin sum, where spins $J_f$ from different $f$ are independent.

The summand of $\sum_{J_f}$ can be written as (up to a factor of $\dim(J_f)$) \cite{hanPI}
\be
\int\rmd g_{ve}\!\sum_{\{M_{ef}\}} \prod_{(v, f)}\left\langle J_f,\g J_f; J_{f},M_{ef}\big| g_{ev}g_{ve^{\prime }}\big| J_f,\g J_f;J_{f},M_{e^{\prime }f}\right\rangle.\label{summand}
\ee
The inner product takes place in the $\Slc$ unitary irrep $\ch^{(J,\g J)}\simeq \oplus_{k=J}^\infty V_k$, where $V_k$ is the irrep of an SU(2) subgroup of $\Slc$. The canonical basis $|J,\g J, J, M\rangle$ is a state in the lowest-level $V_{k=J}$, where $m$ is the magnetic quantum number. Each of the inner product associates to a triangle $f$ and a vertex $v$ of $f$. $e,e'$ label the edges adjacent to $v$.

We pick a $g_{ve}$ and make a change of variable $g_{ve}\to g_{ve} h_e$, $h_e\in \Su$, followed by an integration $\int_{\Su}\rmd h_e$. The operation doesn't change the value of Eq.\Ref{summand} because of the normalization of the Haar measure $\rmd h_e$ on SU(2). $\rmd (g_{ve}h_e)=\rmd g_{ve}$ because $\rmd g_{ve}$ is a Haar measure on $\Slc$. Thus the integral $\int_{\Su}\rmd h_e$ operates as follows:
\be
\int_{\Su}\rmd h_e \prod_{f,e\subset f} h_e|J_f,\g J_f; J_{f},M_{ef}\rangle.
\ee
It only affects 4 states $|J_f,\g J_f; J_{f},M_{ef}\rangle$ whose $f$ contains the edge $e$. $h_e$ leaves $V_{j}$ invariant. $h_e|J_f,\g J_f; J_{f},M_{ef}\rangle$ is essentially the same as $h_e|J_{f},M_{ef}\rangle$. The integral $\int_{\Su}\rmd h_e \prod_{f,e\subset f} h_e$ is a projector onto the invariant subspace of the tensor product $V_{J_1}\otimes\cdots\otimes V_{J_4}$. If 4 $J_f$'s only give a trivial invariant subspace, the above integral vanishes identically for all $M_{ef}$. Indeed we consider the matrix element
\begin{widetext}
\be
\int_{\Su}\rmd h \prod_{i=1}^4 \langle J_i,N_i|h|J_{i},M_{i}\rangle
&=&\sum_{J=|J_1-J_2|}^{J_1+J_2}\sum_{K=|J_3-J_4|}^{J_3+J_4}C^{J_1,J_2;J}_{N_1,N_2;N_1+N_2}C^{J_1,J_2;J}_{M_1,M_2;M_1+M_2}C^{J_3,J_4;K}_{N_3,N_4;N_3+N_4}C^{J_3,J_4;K}_{M_3,M_4;M_3+M_4}\nonumber\\
&\times&\int_{\Su}\rmd h \langle J,N_1+N_2|h|J,M_1+M_{2}\rangle\langle K,N_3+N_4|h|K,M_3+M_{4}\rangle\nonumber\\
&=&\sum_{J=|J_1-J_2|}^{J_1+J_2}\sum_{K=|J_3-J_4|}^{J_3+J_4}C^{J_1,J_2;J}_{N_1,N_2;N_1+N_2}C^{J_1,J_2;J}_{M_1,M_2;M_1+M_2}C^{J_3,J_4;K}_{N_3,N_4;N_3+N_4}C^{J_3,J_4;K}_{M_3,M_4;M_3+M_4}\nonumber\\
&\times&\sum_{\tilde{J}=|J-K|}^{J+K}C^{J,K,\tilde{J}}_{N_1+N_2,N_3+N_4,\tilde{N}}C^{J,K,\tilde{J}}_{M_1+M_2,M_3+M_4,\tilde{M}}\int_{\Su}\rmd h \langle \tilde{J},\tilde{N}|h|\tilde{J},\tilde{M}\rangle
\ee
where the last integral gives $\int_{\Su}\rmd h \langle \tilde{J},\tilde{N}|h|\tilde{J},\tilde{M}\rangle=
\delta_{\tilde{J},0}\delta_{\tilde{M},0}\delta_{\tilde{N},0}$. It constrains 
\be
J=K,\quad N_1+N_2+N_3+N_4=0,\quad M_1+M_2+M_3+M_4=0
\ee
$(J_1,J_2,J)$, $(J_3,J_4,K)$ satisfying triangle inequality and $J=K$ implies there is a nontrivial invariant subspace. If $J\neq K$ the integral vanishes identically.

Note that in the above we have used the product formula of representation matrices: 
\be
\langle J_1,N_1|h|J_{1},M_{1}\rangle\langle J_2,N_2|h_e|J_{2},M_{2}\rangle&=&\sum_{J=|J_1-J_2|}^{J_1+J_2}C^{J_1,J_2;J}_{N_1,N_2;N_1+N_2}C^{J_1,J_2;J}_{M_1,M_2;M_1+M_2}\langle J,N_1+N_2|h|J,M_1+M_{2}\rangle\nonumber\\
\langle J_3,N_3|h|J_{3},M_{3}\rangle\langle J_4,N_4|h_e|J_{4},M_{4}\rangle&=&\sum_{K=|J_3-J_4|}^{J_3+J_4}C^{J_3,J_4;K}_{N_3,N_4;N_3+N_4}C^{J_3,J_4;K}_{M_3,M_4;M_3+M_4}\langle K,N_3+N_4|h|K,M_3+M_{4}\rangle\nonumber
\ee
where $C^{J_1,J_2;J}_{M_1,M_2;M_1+M_2}$ is the Clebsch-Gordan coefficient.

\end{widetext}

We can understand the spin sum $\sum_{J_f}$ as a sum over independent spins, while the integral in the summand imposes the constraint that $J_f$'s should give nontrivial invariant subspace for 4 $f$'s sharing the same edge $e$. For spins in $\sum_{J_f}$ which doesn't satisfy the constraint, their contributions vanish. 

What we have done in the main text is simply interchanging the spin sum and integral. Schematically,
\be
\sum_{J}\dim(J)\int\rmd g\rmd z\ e^{S[J,g,z]}=\int\rmd g\rmd z\sum_{J}\dim(J)\,e^{S[J,g,z]}.
\ee
This interchange can be justified by understanding $\sum_{J}$ as a finite sum, where a large-$J$ cut-off is imposed. The cut-off may relate to the cosmological constant. As another independent justification of interchanging spin sum and integral, we focus on the compact neighborhood $\sn_{Regge}$ in the submanifold $\sm_{Regge}$ in the main discussion. $\sn_{Regge}$ only has finitely many spins (representatives). The spin sum in transverse directions has been regularized by a Gaussian weight with regulator $\delta$, which exponentially decays at infinity as $\delta\neq 0$. It qualifies to interchange the transverse spin sum with the integral.

\section{Transverse Lattice Plane}\label{lattice}

The lattice of all spins $\Fl_J$ is isomorphic to $\Z^{N_f}$, where a lattice basis can be chosen to be $\vec{b}^I=(b^I_f)_{f=1}^{N_f}$ ($I=1,\cdots,N_f$) where $b^I_f=\delta^I_f$. We define a square matrix $B=(\vec{b}^1,\cdots,\vec{b}^{N_f})$ and denote $\Fl_J\simeq\Z^{N_f}=\Fl(B)$. Obviously $B$ is an identity matrix.

A unimodular matrix is a matrix $U\in \Z^{N_f}\times\Z^{N_f}$ such that $\det U=1$. Unimordular matrices relate equivalent lattice bases. Namely columns of $B'=BU$ is a basis of $\Z^{N_f}$ equivalent to the standard basis $\vec{b}^I$. Here $B'$ is simply $U$ since $B$ is an identity matrix. Thus columns of $B'$ give a basis of $\Z^{N_f}$ if and only if it is unimodular. 

The basis from $B'$ is obtained from $B$ via the following operations on columns (unimodular transformation): (1) adding the $I$-th column $n$ times to the $J$-th column, (2) interchanging two columns, and (3) flipping the sign of a column. 

The local neighborhood $\sn_{Regge}\subset\sm_{Regge}$ can be viewed approximately as a $(N_f-M)$ dimensional plane in $\R^{N_f}$. Among the original basis vectors $\vec{b}^I$, there should have been a set of vectors $\vec{b}^K$, say $K=1,\cdots,M_0$, $M_0\leq M$, transverse nicely to $\sn_{Regge}$, i.e. $\vec{b}^K$ doesn't close to any tangent vector of $\sn_{Regge}$. If $M_0<M$ and $\vec{b}^J$ is relatively close to a tangent vector of $\sn_{Regge}$, $\vec{b}^J$ can be improved by the unimodular transformation $\vec{b}^J\to \vec{b}^J+\sum_{K=1}^{M_0}n_K\vec{b}^K$, $n_K\in\Z$, which gives a better transverse lattice vector. Iterating this procedure leads to $M$ transverse lattice vector, while the procedure corresponds to a unimodular matrix $U$, such that $B'=BU$ gives a new basis as its columns. The new basis contains $M$ transverse basis vectors $\hat{e}^i$ which span $\Fl_{NR}$.

\section{Poisson Resummation and Euler-Maclaurin formula}\label{EMF}

In the discussion of the spin sum in Section \ref{2Reg}, we have used the Poisson resummation formula to carry out the sum over $t$. The sum is of the following type
\be
\sum_{t\in\Z}e^{-\delta t^2+t\Phi}=\sum_{k\in\Z}\int_\R e^{-\delta t^2+t(\Phi+2\pi i k)}\rmd t
\ee
where the integral for each $k$ are computed explicitly. 

However the sum can also be studied by the asymptotic expansion using Euler-Maclaurin formula
\be
\sum_{i=m}^n f(i) &=&
    \int^n_m f(x)\,\rmd x + \frac{f(n) + f(m)}{2}\nonumber\\
&+&
    \sum_{k=1}^{\lfloor p/2\rfloor} \frac{B_{2k}}{(2k)!} (f^{(2k - 1)}(n) - f^{(2k - 1)}(m)) + R\label{EMapp}
\ee
where $B_{2k}$ is the $k$-th Bernoulli number. The error term $R$ depends on $n,m, p$ and $f$
\be
R = (-1)^{p+1}\int_m^n f^{(p)}(x) {P_{p}(x) \over p!}\,\rmd x,
\ee 
where $P_p(x)$ is the periodic Bernoulli function. $R$ satisfies the following bound
\be
\left|R\right|\leq\frac{2 \zeta (p)}{(2\pi)^{p}}\int_m^n\left|f^{(p)}(x)\right|
\ee

Let $f(t)=e^{-\delta t^2+t\Phi}$ (exponentially decay at $t\to\pm\infty$), we obtain
\be
\sum_{t\in\Z}e^{-\delta t^2+t\Phi}=\int_\R e^{-\delta t^2+t\Phi}\rmd t+R
\ee
The first term is the same as the $k=0$ term in the Poisson resummation. However since $f^{(p)}\sim \Phi^pe^{-\delta t^2+t\Phi} $, the error term $R$ is not negligible unless $\Phi$ is small. $R$ essentially collects the sum of all $k\neq 0$ contribution in the Poisson resummation.

Viewing $\sum_{t\in\Z}e^{-\delta t^2+t\Phi}$ is a function of $\Phi$, it is clear that replacing sum by integral is only a local approximation of the function (the meaning of asymptotic expansion). $\sum_{t\in\Z}e^{-\delta t^2+t\Phi}$ is periodic in $\Phi\to \Phi+2\pi i$, while $\int e^{-\delta t^2+t\Phi}\rmd t$ breaks the periodicity. The periodicity is not manifest in the Euler-Maclaurin expansion, but is manifest in the Poisson resummation formula.

The small $\Phi$ relates to the small $\g\eps_f$ in Section \ref{RESDA}. Thus the result with $k=0$ in Section \ref{RESDA} can be reproduced by using the Euler-Maclaurin expansion in the regime where $R$ is negligible. The ER regime essentially requires $\sum_{t\in\Z}e^{-\delta t^2+t\Phi}$ can be approximated by $\int e^{-\delta t^2+t\Phi}\rmd t$

Similarly when one consider the large-$J$ spin sum in spinfoam amplitude, one would like to rescale $J_f=\l j_f$ where $\Delta j_f=\frac{1}{2\l}$ ($\l\gg1$) and understand the spin sum as the Riemann sum, i.e. schematically 
\be
\sum_j e^{\l j F}=2\l \sum_j \Delta j\, e^{\l j F}\sim 2\l \int\rmd j\,  e^{\l j F}=2 \int\rmd J\,  e^{J F}\nonumber
\ee
However because of the Euler-Maclaurin expansion Eq.\Ref{EMapp}, we know that the above approximation may valid only in the regime of small $F$. In general the error terms are not negligible. It can also be seen in the Euler-Maclaurin expansion of $\sum_j \Delta j\, f(j)$ where $f(j)=e^{\l j F}$. The $\l^{-n}$ correction involves the $n$-th derivative $f^{(n)}(j)=\l^n F^n e^{\l j F}$ which cancels $\l^{-n}$.

In the discussion of the variational principle of Regge action in Section \ref{RESDA}. We have implicitly used the Euler-Maclaurin expansion for Eq.\Ref{Zregge}
\be
\sum_{\ell} e^{\frac{i}{\ell_P^2}S_{Regge}[\ell]+\cdots}= \int\rmd{\ell}\,  e^{\frac{i}{\ell_P^2}S_{Regge}[\ell]+\cdots}+\text{error terms}
\ee
In general the error terms are not negligible as far as the full amplitude is concerned. However as far as the equation of motion is concerned, the variational principle is applied to the first term, whose dominant contribution comes from solutions of the Regge equation.

\section{Action and Angles in Euclidean EPRL amplitude }\label{Eucl}

Consider an internal dual face $f$, at each large-$J$ critical point (of a globally oriented nondegenerate geometry) in the Euclidean spinfoam amplitude, the loop holonomy along $\partial f$ made by $g_{ve}^\pm$'s is written as
\be
G_f^\pm(v)\equiv g_{ve}^{\pm}g_{ev_{k}}^{\pm }g_{v_{k}e_{k}}^{\pm }\cdots g_{e_{1}v}^{\pm }=\exp \left( i\Phi _{f}^{\pm }\hat{X}_{f}^{\pm}(v)\right) 
\ee
where $\hat{X}_{f}(v)=(\hat{X}_{f}^{+}(v),\hat{X}_{f}^{-}(v))$ is the normalized bivector along the triangle $f$. $\Phi _{f}^{\pm }=\sum_{v}\phi^\pm_{eve'}$ where $\phi^\pm_{eve'}$ within each 4-simplex satisfies \cite{semiclassicalEu,HZ1}
\be
\phi^+_{eve'}-\phi^-_{eve'}=\mu(v)\,\Theta_{f}(v)\in[-\pi,\pi].\label{phiphi}
\ee
$\mu(v)$ relates to the orientation of the 4-simplex $v$, which we set to be globally $\mu(v)=-1$ for globally oriented spacetime geometries. 

The action contributed by $f$ evaluated at the critical point reads \cite{CFsemiclassical,HZ1},
\be
S_{f} =\sum_{\pm }2iJ_{f}^{\pm }\Phi _{f}^{\pm }=iJ_f\lt(\Phi_f^++\Phi_f^-\rt)+i\g J_f\lt(\Phi_f^+-\Phi_f^-\rt)
\ee
Each $\Phi _{f}^{\pm }$ is defined modulo $2\pi$: $\Phi _{f}^{\pm }\sim\Phi _{f}^{\pm }+2\pi$. So $\Phi_f^+\pm\Phi_f^-\sim \Phi_f^+\pm\Phi_f^-+4\pi$. However simultaneous transformations $\Phi_f^+\pm\Phi_f^-\to \Phi_f^+\pm\Phi_f^-+2\pi$ doesn't change $e^{S_f}$ since $(1+\g)j_f\in\Z$. We can set the following range of angles:
\be
\Phi_f^++\Phi_f^-\in[-2\pi,2\pi],\quad \Phi_f^+-\Phi_f^-\in[-\pi,\pi].\label{B3}
\ee

Eq.\Ref{phiphi} implies $\Phi_f^+-\Phi_f^-= -\sum_{v\in f}\Theta_f(v)$ mod $4\pi$. But simultaneous transformations can give
\be
\Phi_f^+-\Phi_f^-=2\pi -\sum_{v\in f}\Theta_f(v)=\eps_f,
\ee
when we set $\eps_f\in[-\pi,\pi]$ to include Regge geometries close to the continuum. $\eps_f\in[-\pi,\pi]$ is made by choosing suitable $\sn_{Regge}$.

On the other hand, $G_f(e)$ represented in the vector representation $\hat{G}_f(e)$ reads \cite{CFsemiclassical,HZ1}:
\be
\hat{G}_f(v)=\exp\left( *\hat{X}_{f}(v)\,\theta_f\right)\exp\lt(\pi\eta_f\hat{X}_{f}(v)\rt).
\ee
where $\eta_f\in\{0,1\}$ labels 2 different types of critical points.

Lifting $\hat{G}_f(v)\in \rm{SO}(4)$ to $(G_f^+(v),G_f^-(v))\in\Su\times\Su$ evaluates $\Phi^{\pm}_f=\frac{1}{2}\lt(\eta_f\pi\pm\theta_f\rt)-k_f\pi,$ where $k_f\in\{0,1\}$ label lift ambiguities.
\be
\eps_f=\Phi_f^+-\Phi_f^-=\theta_f ,\nonumber\\
 \Phi_f^++\Phi_f^-=\pi\eta_f-2k_f \pi
\ee
Eq.\Ref{B3} implies 
\be
\eta_f-2k_f\equiv n_f\in\{-1,0,1\}.
\ee
There is canonical lift with $k_f=0$ corresponding the lift of SO(4) spin connection to $\Su\times \Su$. $\eta_f=k_f=0$ indeed corresponds to a critical solution, which can be constructed by the Regge geometry with the canonical lift\footnote{The construction of critical solutions from arbitrary Regge geometries can be done locally in each 4-simplex as explained in \cite{semiclassicalEu}, see also \cite{hanPI} in Lorentzian signature.}. Other lifts $k^\pm\neq0$ and $\eta_f\neq0$ may corresponds to different critical solutions\footnote{One should also take into account the gauge invariance $g_{ve}^\pm\to\kappa_{ve} g_{ve}^\pm$ ($\kappa_{ve}=\pm1$) of spinfoam action, which removes some lift ambiguities.}.


The action is expressed as
\be
S_f=iJ_f\lt[\g\eps_f+n_f\pi\rt].
\ee
Therefore repeating the analysis in Section \ref{RESDA} leads to the replacement 
\be
\g\eps_f\to \g\eps_f+n_f\pi
\ee
in Eqs.\Ref{ER} and \Ref{smalleps}. After the replacement Eq.\Ref{smalleps} gives disjoint sectors of geometries whose $\g\eps_f$ are close to $-n_f\pi$. The only sector having geometries approximating the continuum is the one with all $n_f=0$. Other sectors are suppressed in the amplitude by suitably choosing $\sn_{Regge}$.


\begin{thebibliography}{72}
\expandafter\ifx\csname natexlab\endcsname\relax\def\natexlab#1{#1}\fi
\expandafter\ifx\csname bibnamefont\endcsname\relax
  \def\bibnamefont#1{#1}\fi
\expandafter\ifx\csname bibfnamefont\endcsname\relax
  \def\bibfnamefont#1{#1}\fi
\expandafter\ifx\csname citenamefont\endcsname\relax
  \def\citenamefont#1{#1}\fi
\expandafter\ifx\csname url\endcsname\relax
  \def\url#1{\texttt{#1}}\fi
\expandafter\ifx\csname urlprefix\endcsname\relax\def\urlprefix{URL }\fi
\providecommand{\bibinfo}[2]{#2}
\providecommand{\eprint}[2][]{\url{#2}}

\bibitem[{\citenamefont{Thiemann}(2007)}]{book}
\bibinfo{author}{\bibfnamefont{T.}~\bibnamefont{Thiemann}},
  \emph{\bibinfo{title}{Modern Canonical Quantum General Relativity}}
  (\bibinfo{publisher}{Cambridge University Press}, \bibinfo{year}{2007}).

\bibitem[{\citenamefont{Han et~al.}(2007)\citenamefont{Han, Huang, and
  Ma}}]{review}
\bibinfo{author}{\bibfnamefont{M.}~\bibnamefont{Han}},
  \bibinfo{author}{\bibfnamefont{W.}~\bibnamefont{Huang}}, \bibnamefont{and}
  \bibinfo{author}{\bibfnamefont{Y.}~\bibnamefont{Ma}},
  \bibinfo{journal}{Int.J.Mod.Phys.} \textbf{\bibinfo{volume}{D16}},
  \bibinfo{pages}{1397} (\bibinfo{year}{2007}), \eprint{gr-qc/0509064}.

\bibitem[{\citenamefont{Ashtekar and Lewandowski}(2004)}]{review1}
\bibinfo{author}{\bibfnamefont{A.}~\bibnamefont{Ashtekar}} \bibnamefont{and}
  \bibinfo{author}{\bibfnamefont{J.}~\bibnamefont{Lewandowski}},
  \bibinfo{journal}{Class.Quant.Grav.} \textbf{\bibinfo{volume}{21}},
  \bibinfo{pages}{R53} (\bibinfo{year}{2004}), \eprint{gr-qc/0404018}.

\bibitem[{\citenamefont{Rovelli and Vidotto}(2014)}]{rovelli2014covariant}
\bibinfo{author}{\bibfnamefont{C.}~\bibnamefont{Rovelli}} \bibnamefont{and}
  \bibinfo{author}{\bibfnamefont{F.}~\bibnamefont{Vidotto}},
  \emph{\bibinfo{title}{Covariant Loop Quantum Gravity: An Elementary
  Introduction to Quantum Gravity and Spinfoam Theory}}, Cambridge Monographs
  on Mathematical Physics (\bibinfo{publisher}{Cambridge University Press},
  \bibinfo{year}{2014}), ISBN \bibinfo{isbn}{9781107069626}.

\bibitem[{\citenamefont{Perez}(2013)}]{Perez2012}
\bibinfo{author}{\bibfnamefont{A.}~\bibnamefont{Perez}},
  \bibinfo{journal}{Living Rev.Rel.} \textbf{\bibinfo{volume}{16}},
  \bibinfo{pages}{3} (\bibinfo{year}{2013}), \eprint{1205.2019}.

\bibitem[{\citenamefont{Conrady and Freidel}(2008)}]{CFsemiclassical}
\bibinfo{author}{\bibfnamefont{F.}~\bibnamefont{Conrady}} \bibnamefont{and}
  \bibinfo{author}{\bibfnamefont{L.}~\bibnamefont{Freidel}},
  \bibinfo{journal}{Phys.Rev.} \textbf{\bibinfo{volume}{D78}},
  \bibinfo{pages}{104023} (\bibinfo{year}{2008}), \eprint{0809.2280}.

\bibitem[{\citenamefont{Barrett et~al.}(2010)\citenamefont{Barrett, Dowdall,
  Fairbairn, Hellmann, and Pereira}}]{semiclassical}
\bibinfo{author}{\bibfnamefont{J.~W.} \bibnamefont{Barrett}},
  \bibinfo{author}{\bibfnamefont{R.}~\bibnamefont{Dowdall}},
  \bibinfo{author}{\bibfnamefont{W.~J.} \bibnamefont{Fairbairn}},
  \bibinfo{author}{\bibfnamefont{F.}~\bibnamefont{Hellmann}}, \bibnamefont{and}
  \bibinfo{author}{\bibfnamefont{R.}~\bibnamefont{Pereira}},
  \bibinfo{journal}{Class.Quant.Grav.} \textbf{\bibinfo{volume}{27}},
  \bibinfo{pages}{165009} (\bibinfo{year}{2010}), \eprint{0907.2440}.

\bibitem[{\citenamefont{Han and Zhang}(2013)}]{HZ}
\bibinfo{author}{\bibfnamefont{M.}~\bibnamefont{Han}} \bibnamefont{and}
  \bibinfo{author}{\bibfnamefont{M.}~\bibnamefont{Zhang}},
  \bibinfo{journal}{Class.Quant.Grav.} \textbf{\bibinfo{volume}{30}},
  \bibinfo{pages}{165012} (\bibinfo{year}{2013}), \eprint{1109.0499}.

\bibitem[{\citenamefont{Haggard
  et~al.}(2015{\natexlab{a}})\citenamefont{Haggard, Han, Kaminski, and
  Riello}}]{HHKR}
\bibinfo{author}{\bibfnamefont{H.~M.} \bibnamefont{Haggard}},
  \bibinfo{author}{\bibfnamefont{M.}~\bibnamefont{Han}},
  \bibinfo{author}{\bibfnamefont{W.}~\bibnamefont{Kaminski}}, \bibnamefont{and}
  \bibinfo{author}{\bibfnamefont{A.}~\bibnamefont{Riello}},
  \bibinfo{journal}{Nucl. Phys.} \textbf{\bibinfo{volume}{B900}},
  \bibinfo{pages}{1} (\bibinfo{year}{2015}{\natexlab{a}}), \eprint{1412.7546}.

\bibitem[{\citenamefont{Kaminski et~al.}(2017)\citenamefont{Kaminski,
  Kisielowski, and Sahlmann}}]{Kaminski:2017eew}
\bibinfo{author}{\bibfnamefont{W.}~\bibnamefont{Kaminski}},
  \bibinfo{author}{\bibfnamefont{M.}~\bibnamefont{Kisielowski}},
  \bibnamefont{and} \bibinfo{author}{\bibfnamefont{H.}~\bibnamefont{Sahlmann}}
  (\bibinfo{year}{2017}), \eprint{1705.02862}.

\bibitem[{\citenamefont{Magliaro and Perini}(2011)}]{claudio1}
\bibinfo{author}{\bibfnamefont{E.}~\bibnamefont{Magliaro}} \bibnamefont{and}
  \bibinfo{author}{\bibfnamefont{C.}~\bibnamefont{Perini}},
  \bibinfo{journal}{Europhys.Lett.} \textbf{\bibinfo{volume}{95}},
  \bibinfo{pages}{30007} (\bibinfo{year}{2011}), \eprint{1108.2258}.

\bibitem[{\citenamefont{Bianchi and Ding}(2012)}]{propagator3}
\bibinfo{author}{\bibfnamefont{E.}~\bibnamefont{Bianchi}} \bibnamefont{and}
  \bibinfo{author}{\bibfnamefont{Y.}~\bibnamefont{Ding}},
  \bibinfo{journal}{Phys.Rev.} \textbf{\bibinfo{volume}{D86}},
  \bibinfo{pages}{104040} (\bibinfo{year}{2012}), \eprint{1109.6538}.

\bibitem[{\citenamefont{Bianchi et~al.}(2009)\citenamefont{Bianchi, Magliaro,
  and Perini}}]{propagator2}
\bibinfo{author}{\bibfnamefont{E.}~\bibnamefont{Bianchi}},
  \bibinfo{author}{\bibfnamefont{E.}~\bibnamefont{Magliaro}}, \bibnamefont{and}
  \bibinfo{author}{\bibfnamefont{C.}~\bibnamefont{Perini}},
  \bibinfo{journal}{Nucl.Phys.} \textbf{\bibinfo{volume}{B822}},
  \bibinfo{pages}{245} (\bibinfo{year}{2009}), \eprint{0905.4082}.

\bibitem[{\citenamefont{Bianchi et~al.}(2006)\citenamefont{Bianchi, Modesto,
  Rovelli, and Speziale}}]{propagator1}
\bibinfo{author}{\bibfnamefont{E.}~\bibnamefont{Bianchi}},
  \bibinfo{author}{\bibfnamefont{L.}~\bibnamefont{Modesto}},
  \bibinfo{author}{\bibfnamefont{C.}~\bibnamefont{Rovelli}}, \bibnamefont{and}
  \bibinfo{author}{\bibfnamefont{S.}~\bibnamefont{Speziale}},
  \bibinfo{journal}{Class.Quant.Grav.} \textbf{\bibinfo{volume}{23}},
  \bibinfo{pages}{6989} (\bibinfo{year}{2006}), \eprint{gr-qc/0604044}.

\bibitem[{\citenamefont{Alesci}(2009)}]{Alesci:2009ys}
\bibinfo{author}{\bibfnamefont{E.}~\bibnamefont{Alesci}}, in
  \emph{\bibinfo{booktitle}{{3rd Stueckelberg Workshop on Relativistic Field
  Theories Pescara, Italy, July 8-18, 2008}}} (\bibinfo{year}{2009}),
  \eprint{0903.4329}.

\bibitem[{\citenamefont{Rovelli and Zhang}(2011)}]{3pt}
\bibinfo{author}{\bibfnamefont{C.}~\bibnamefont{Rovelli}} \bibnamefont{and}
  \bibinfo{author}{\bibfnamefont{M.}~\bibnamefont{Zhang}},
  \bibinfo{journal}{Class.Quant.Grav.} \textbf{\bibinfo{volume}{28}},
  \bibinfo{pages}{175010} (\bibinfo{year}{2011}), \eprint{1105.0566}.

\bibitem[{\citenamefont{Bonzom}(2009)}]{flatness}
\bibinfo{author}{\bibfnamefont{V.}~\bibnamefont{Bonzom}},
  \bibinfo{journal}{Phys. Rev.} \textbf{\bibinfo{volume}{D80}},
  \bibinfo{pages}{064028} (\bibinfo{year}{2009}), \eprint{0905.1501}.

\bibitem[{\citenamefont{Hellmann and Kaminski}(2013)}]{frankflat}
\bibinfo{author}{\bibfnamefont{F.}~\bibnamefont{Hellmann}} \bibnamefont{and}
  \bibinfo{author}{\bibfnamefont{W.}~\bibnamefont{Kaminski}},
  \bibinfo{journal}{JHEP} \textbf{\bibinfo{volume}{10}}, \bibinfo{pages}{165}
  (\bibinfo{year}{2013}), \eprint{1307.1679}.

\bibitem[{\citenamefont{Perini}(2012)}]{Perini:2012nd}
\bibinfo{author}{\bibfnamefont{C.}~\bibnamefont{Perini}}
  (\bibinfo{year}{2012}), \eprint{1211.4807}.

\bibitem[{\citenamefont{Han and Zhang}(2016)}]{Han:2016fgh}
\bibinfo{author}{\bibfnamefont{M.}~\bibnamefont{Han}} \bibnamefont{and}
  \bibinfo{author}{\bibfnamefont{M.}~\bibnamefont{Zhang}},
  \bibinfo{journal}{Phys. Rev.} \textbf{\bibinfo{volume}{D94}},
  \bibinfo{pages}{104075} (\bibinfo{year}{2016}), \eprint{1606.02826}.

\bibitem[{\citenamefont{Freidel}(2014)}]{freideltalk}
\bibinfo{author}{\bibfnamefont{L.}~\bibnamefont{Freidel}},
  \bibinfo{journal}{ILQGS talk}  (\bibinfo{year}{2014}).

\bibitem[{\citenamefont{Feinberg et~al.}(1984)\citenamefont{Feinberg,
  Friedberg, Lee, and Ren}}]{FFLR}
\bibinfo{author}{\bibfnamefont{G.}~\bibnamefont{Feinberg}},
  \bibinfo{author}{\bibfnamefont{R.}~\bibnamefont{Friedberg}},
  \bibinfo{author}{\bibfnamefont{T.}~\bibnamefont{Lee}}, \bibnamefont{and}
  \bibinfo{author}{\bibfnamefont{H.}~\bibnamefont{Ren}},
  \bibinfo{journal}{Nucl.Phys.} \textbf{\bibinfo{volume}{B245}},
  \bibinfo{pages}{343} (\bibinfo{year}{1984}).

\bibitem[{\citenamefont{Barrett and Parker}(1994)}]{BARRETT1994107}
\bibinfo{author}{\bibfnamefont{J.}~\bibnamefont{Barrett}} \bibnamefont{and}
  \bibinfo{author}{\bibfnamefont{P.}~\bibnamefont{Parker}},
  \bibinfo{journal}{Journal of Approximation Theory}
  \textbf{\bibinfo{volume}{76}}, \bibinfo{pages}{107 } (\bibinfo{year}{1994}),
  ISSN \bibinfo{issn}{0021-9045}.

\bibitem[{\citenamefont{Han and Zhang}(2012)}]{HZ1}
\bibinfo{author}{\bibfnamefont{M.}~\bibnamefont{Han}} \bibnamefont{and}
  \bibinfo{author}{\bibfnamefont{M.}~\bibnamefont{Zhang}},
  \bibinfo{journal}{Class.Quant.Grav.} \textbf{\bibinfo{volume}{29}},
  \bibinfo{pages}{165004} (\bibinfo{year}{2012}), \eprint{1109.0500}.

\bibitem[{\citenamefont{Han and Krajewski}(2014)}]{hanPI}
\bibinfo{author}{\bibfnamefont{M.}~\bibnamefont{Han}} \bibnamefont{and}
  \bibinfo{author}{\bibfnamefont{T.}~\bibnamefont{Krajewski}},
  \bibinfo{journal}{Class.Quant.Grav.} \textbf{\bibinfo{volume}{31}},
  \bibinfo{pages}{015009} (\bibinfo{year}{2014}), \eprint{1304.5626}.

\bibitem[{\citenamefont{Williams and Tuckey}(1992)}]{Williams:1991cd}
\bibinfo{author}{\bibfnamefont{R.~M.} \bibnamefont{Williams}} \bibnamefont{and}
  \bibinfo{author}{\bibfnamefont{P.~A.} \bibnamefont{Tuckey}},
  \bibinfo{journal}{Class. Quant. Grav.} \textbf{\bibinfo{volume}{9}},
  \bibinfo{pages}{1409} (\bibinfo{year}{1992}).

\bibitem[{\citenamefont{Gentle}(2002)}]{Gentle:2002ux}
\bibinfo{author}{\bibfnamefont{A.~P.} \bibnamefont{Gentle}},
  \bibinfo{journal}{Gen. Rel. Grav.} \textbf{\bibinfo{volume}{34}},
  \bibinfo{pages}{1701} (\bibinfo{year}{2002}), \eprint{gr-qc/0408006}.

\bibitem[{\citenamefont{Barrett and Williams}(1988)}]{0264-9381-5-12-007}
\bibinfo{author}{\bibfnamefont{J.~W.} \bibnamefont{Barrett}} \bibnamefont{and}
  \bibinfo{author}{\bibfnamefont{R.~M.} \bibnamefont{Williams}},
  \bibinfo{journal}{Classical and Quantum Gravity}
  \textbf{\bibinfo{volume}{5}}, \bibinfo{pages}{1543} (\bibinfo{year}{1988}).

\bibitem[{\citenamefont{Barrett}(1988{\natexlab{a}})}]{0264-9381-5-9-004}
\bibinfo{author}{\bibfnamefont{J.~W.} \bibnamefont{Barrett}},
  \bibinfo{journal}{Classical and Quantum Gravity}
  \textbf{\bibinfo{volume}{5}}, \bibinfo{pages}{1187}
  (\bibinfo{year}{1988}{\natexlab{a}}).

\bibitem[{\citenamefont{Christiansen}(2011)}]{Christiansen2011}
\bibinfo{author}{\bibfnamefont{S.~H.} \bibnamefont{Christiansen}},
  \bibinfo{journal}{Numerische Mathematik} \textbf{\bibinfo{volume}{119}},
  \bibinfo{pages}{613} (\bibinfo{year}{2011}), ISSN \bibinfo{issn}{0945-3245}.

\bibitem[{\citenamefont{Gentle}(2013)}]{Gentle:2012tc}
\bibinfo{author}{\bibfnamefont{A.~P.} \bibnamefont{Gentle}},
  \bibinfo{journal}{Class. Quant. Grav.} \textbf{\bibinfo{volume}{30}},
  \bibinfo{pages}{085004} (\bibinfo{year}{2013}), \eprint{1208.1502}.

\bibitem[{\citenamefont{Gentle and Miller}(1998)}]{Gentle:1997df}
\bibinfo{author}{\bibfnamefont{A.~P.} \bibnamefont{Gentle}} \bibnamefont{and}
  \bibinfo{author}{\bibfnamefont{W.~A.} \bibnamefont{Miller}},
  \bibinfo{journal}{Class. Quant. Grav.} \textbf{\bibinfo{volume}{15}},
  \bibinfo{pages}{389} (\bibinfo{year}{1998}), \eprint{gr-qc/9706034}.

\bibitem[{\citenamefont{Gentle et~al.}(1999)\citenamefont{Gentle, Holz, Miller,
  and Wheeler}}]{Gentle:1998qg}
\bibinfo{author}{\bibfnamefont{A.~P.} \bibnamefont{Gentle}},
  \bibinfo{author}{\bibfnamefont{D.~E.} \bibnamefont{Holz}},
  \bibinfo{author}{\bibfnamefont{W.~A.} \bibnamefont{Miller}},
  \bibnamefont{and} \bibinfo{author}{\bibfnamefont{J.~A.}
  \bibnamefont{Wheeler}}, \bibinfo{journal}{Class. Quant. Grav.}
  \textbf{\bibinfo{volume}{16}}, \bibinfo{pages}{1979} (\bibinfo{year}{1999}),
  \eprint{gr-qc/9812057}.

\bibitem[{\citenamefont{Liu and Williams}(2016)}]{Liu:2015gpa}
\bibinfo{author}{\bibfnamefont{R.~G.} \bibnamefont{Liu}} \bibnamefont{and}
  \bibinfo{author}{\bibfnamefont{R.~M.} \bibnamefont{Williams}},
  \bibinfo{journal}{Phys. Rev.} \textbf{\bibinfo{volume}{D93}},
  \bibinfo{pages}{024032} (\bibinfo{year}{2016}), \eprint{1501.07614}.

\bibitem[{\citenamefont{Miller}(1995)}]{Miller:1995gz}
\bibinfo{author}{\bibfnamefont{M.~A.} \bibnamefont{Miller}},
  \bibinfo{journal}{Class. Quant. Grav.} \textbf{\bibinfo{volume}{12}},
  \bibinfo{pages}{3037} (\bibinfo{year}{1995}), \eprint{gr-qc/9502044}.

\bibitem[{\citenamefont{Gentle et~al.}(2009)\citenamefont{Gentle, Kheyfets,
  McDonald, and Miller}}]{Gentle:2008fy}
\bibinfo{author}{\bibfnamefont{A.~P.} \bibnamefont{Gentle}},
  \bibinfo{author}{\bibfnamefont{A.}~\bibnamefont{Kheyfets}},
  \bibinfo{author}{\bibfnamefont{J.~R.} \bibnamefont{McDonald}},
  \bibnamefont{and} \bibinfo{author}{\bibfnamefont{W.~A.}
  \bibnamefont{Miller}}, \bibinfo{journal}{Class. Quant. Grav.}
  \textbf{\bibinfo{volume}{26}}, \bibinfo{pages}{015005}
  (\bibinfo{year}{2009}), \eprint{0807.3041}.

\bibitem[{\citenamefont{Han}(2014{\natexlab{a}})}]{lowE}
\bibinfo{author}{\bibfnamefont{M.}~\bibnamefont{Han}},
  \bibinfo{journal}{Phys.Rev.} \textbf{\bibinfo{volume}{D89}},
  \bibinfo{pages}{124001} (\bibinfo{year}{2014}{\natexlab{a}}),
  \eprint{1308.4063}.

\bibitem[{\citenamefont{Han}(2014{\natexlab{b}})}]{LowE1}
\bibinfo{author}{\bibfnamefont{M.}~\bibnamefont{Han}},
  \bibinfo{journal}{Class.Quant.Grav.} \textbf{\bibinfo{volume}{31}},
  \bibinfo{pages}{015004} (\bibinfo{year}{2014}{\natexlab{b}}),
  \eprint{1304.5627}.

\bibitem[{\citenamefont{Han}(2013)}]{lowE2}
\bibinfo{author}{\bibfnamefont{M.}~\bibnamefont{Han}},
  \bibinfo{journal}{Phys.Rev.} \textbf{\bibinfo{volume}{D88}},
  \bibinfo{pages}{044051} (\bibinfo{year}{2013}), \eprint{1304.5628}.

\bibitem[{\citenamefont{Cheeger et~al.}(1984)\citenamefont{Cheeger, Muller, and
  Schrader}}]{cheeger1984}
\bibinfo{author}{\bibfnamefont{J.}~\bibnamefont{Cheeger}},
  \bibinfo{author}{\bibfnamefont{W.}~\bibnamefont{Muller}}, \bibnamefont{and}
  \bibinfo{author}{\bibfnamefont{R.}~\bibnamefont{Schrader}},
  \bibinfo{journal}{Comm. Math. Phys.} \textbf{\bibinfo{volume}{92}},
  \bibinfo{pages}{405} (\bibinfo{year}{1984}).

\bibitem[{\citenamefont{Engle et~al.}(2008)\citenamefont{Engle, Livine,
  Pereira, and Rovelli}}]{EPRL}
\bibinfo{author}{\bibfnamefont{J.}~\bibnamefont{Engle}},
  \bibinfo{author}{\bibfnamefont{E.}~\bibnamefont{Livine}},
  \bibinfo{author}{\bibfnamefont{R.}~\bibnamefont{Pereira}}, \bibnamefont{and}
  \bibinfo{author}{\bibfnamefont{C.}~\bibnamefont{Rovelli}},
  \bibinfo{journal}{Nucl.Phys.} \textbf{\bibinfo{volume}{B799}},
  \bibinfo{pages}{136} (\bibinfo{year}{2008}), \eprint{0711.0146}.

\bibitem[{\citenamefont{Freidel and Krasnov}(2008)}]{FK}
\bibinfo{author}{\bibfnamefont{L.}~\bibnamefont{Freidel}} \bibnamefont{and}
  \bibinfo{author}{\bibfnamefont{K.}~\bibnamefont{Krasnov}},
  \bibinfo{journal}{Class.Quant.Grav.} \textbf{\bibinfo{volume}{25}},
  \bibinfo{pages}{125018} (\bibinfo{year}{2008}), \eprint{0708.1595}.

\bibitem[{\citenamefont{Conrady and Hnybida}(2010)}]{Conrady:2010kc}
\bibinfo{author}{\bibfnamefont{F.}~\bibnamefont{Conrady}} \bibnamefont{and}
  \bibinfo{author}{\bibfnamefont{J.}~\bibnamefont{Hnybida}},
  \bibinfo{journal}{Class. Quant. Grav.} \textbf{\bibinfo{volume}{27}},
  \bibinfo{pages}{185011} (\bibinfo{year}{2010}), \eprint{1002.1959}.

\bibitem[{\citenamefont{Barrett et~al.}(2009)\citenamefont{Barrett, Dowdall,
  Fairbairn, Gomes, and Hellmann}}]{semiclassicalEu}
\bibinfo{author}{\bibfnamefont{J.~W.} \bibnamefont{Barrett}},
  \bibinfo{author}{\bibfnamefont{R.}~\bibnamefont{Dowdall}},
  \bibinfo{author}{\bibfnamefont{W.~J.} \bibnamefont{Fairbairn}},
  \bibinfo{author}{\bibfnamefont{H.}~\bibnamefont{Gomes}}, \bibnamefont{and}
  \bibinfo{author}{\bibfnamefont{F.}~\bibnamefont{Hellmann}},
  \bibinfo{journal}{J.Math.Phys.} \textbf{\bibinfo{volume}{50}},
  \bibinfo{pages}{112504} (\bibinfo{year}{2009}), \eprint{0902.1170}.

\bibitem[{\citenamefont{H\"{o}rmander}(1998)}]{stationaryphase}
\bibinfo{author}{\bibfnamefont{L.}~\bibnamefont{H\"{o}rmander}},
  \emph{\bibinfo{title}{The Analysis of Linear Partial Differential Operators
  I}}, vol. \bibinfo{volume}{256} of \emph{\bibinfo{series}{Grundlehren der
  mathematischen Wissenschaften}} (\bibinfo{publisher}{Springer Berlin
  Heidelberg}, \bibinfo{address}{Berlin, Heidelberg}, \bibinfo{year}{1998}),
  ISBN \bibinfo{isbn}{978-3-642-96752-8}.

\bibitem[{\citenamefont{Magliaro and Perini}(2013)}]{claudio}
\bibinfo{author}{\bibfnamefont{E.}~\bibnamefont{Magliaro}} \bibnamefont{and}
  \bibinfo{author}{\bibfnamefont{C.}~\bibnamefont{Perini}},
  \bibinfo{journal}{Int.J.Mod.Phys.} \textbf{\bibinfo{volume}{D22}},
  \bibinfo{pages}{1} (\bibinfo{year}{2013}), \eprint{1105.0216}.

\bibitem[{\citenamefont{Barrett}(1988{\natexlab{b}})}]{BarrettLinRegge}
\bibinfo{author}{\bibfnamefont{J.~W.} \bibnamefont{Barrett}},
  \bibinfo{journal}{Classical and Quantum Gravity}
  \textbf{\bibinfo{volume}{5}}, \bibinfo{pages}{1187}
  (\bibinfo{year}{1988}{\natexlab{b}}).

\bibitem[{\citenamefont{Han and Hung}(2017)}]{HanHung}
\bibinfo{author}{\bibfnamefont{M.}~\bibnamefont{Han}} \bibnamefont{and}
  \bibinfo{author}{\bibfnamefont{L.-Y.} \bibnamefont{Hung}},
  \bibinfo{journal}{Phys. Rev.} \textbf{\bibinfo{volume}{D95}},
  \bibinfo{pages}{024011} (\bibinfo{year}{2017}), \eprint{1610.02134}.

\bibitem[{\citenamefont{Lipatov}(1977)}]{Lipatov}
\bibinfo{author}{\bibfnamefont{L.}~\bibnamefont{Lipatov}},
  \bibinfo{journal}{J.E.T.P. Lett.} \textbf{\bibinfo{volume}{25}},
  \bibinfo{pages}{116} (\bibinfo{year}{1977}).

\bibitem[{\citenamefont{Spencer}(1980)}]{Spencer1980}
\bibinfo{author}{\bibfnamefont{T.}~\bibnamefont{Spencer}},
  \bibinfo{journal}{Comm. Math. Phys.} \textbf{\bibinfo{volume}{74}},
  \bibinfo{pages}{273} (\bibinfo{year}{1980}), ISSN \bibinfo{issn}{1432-0916}.

\bibitem[{\citenamefont{David et~al.}(1988)\citenamefont{David, Feldman, and
  Rivasseau}}]{david1988}
\bibinfo{author}{\bibfnamefont{F.}~\bibnamefont{David}},
  \bibinfo{author}{\bibfnamefont{J.}~\bibnamefont{Feldman}}, \bibnamefont{and}
  \bibinfo{author}{\bibfnamefont{V.}~\bibnamefont{Rivasseau}},
  \bibinfo{journal}{Comm. Math. Phys.} \textbf{\bibinfo{volume}{116}},
  \bibinfo{pages}{215} (\bibinfo{year}{1988}).

\bibitem[{\citenamefont{Christodoulou et~al.}(2012)\citenamefont{Christodoulou,
  Riello, and Rovelli}}]{Christodoulou:2012sm}
\bibinfo{author}{\bibfnamefont{M.}~\bibnamefont{Christodoulou}},
  \bibinfo{author}{\bibfnamefont{A.}~\bibnamefont{Riello}}, \bibnamefont{and}
  \bibinfo{author}{\bibfnamefont{C.}~\bibnamefont{Rovelli}},
  \bibinfo{journal}{Int. J. Mod. Phys.} \textbf{\bibinfo{volume}{D21}},
  \bibinfo{pages}{1242014} (\bibinfo{year}{2012}), \eprint{1206.3903}.

\bibitem[{\citenamefont{Engle et~al.}(2016)\citenamefont{Engle, Vilensky, and
  Zipfel}}]{Engle:2015zqa}
\bibinfo{author}{\bibfnamefont{J.}~\bibnamefont{Engle}},
  \bibinfo{author}{\bibfnamefont{I.}~\bibnamefont{Vilensky}}, \bibnamefont{and}
  \bibinfo{author}{\bibfnamefont{A.}~\bibnamefont{Zipfel}},
  \bibinfo{journal}{Phys. Rev.} \textbf{\bibinfo{volume}{D94}},
  \bibinfo{pages}{064025} (\bibinfo{year}{2016}), \eprint{1505.06683}.

\bibitem[{\citenamefont{Haggard
  et~al.}(2016{\natexlab{a}})\citenamefont{Haggard, Han, Kaminski, and
  Riello}}]{HHKRshort}
\bibinfo{author}{\bibfnamefont{H.~M.} \bibnamefont{Haggard}},
  \bibinfo{author}{\bibfnamefont{M.}~\bibnamefont{Han}},
  \bibinfo{author}{\bibfnamefont{W.}~\bibnamefont{Kaminski}}, \bibnamefont{and}
  \bibinfo{author}{\bibfnamefont{A.}~\bibnamefont{Riello}},
  \bibinfo{journal}{Phys. Lett.} \textbf{\bibinfo{volume}{B752}},
  \bibinfo{pages}{258} (\bibinfo{year}{2016}{\natexlab{a}}),
  \eprint{1509.00458}.

\bibitem[{\citenamefont{Haggard
  et~al.}(2015{\natexlab{b}})\citenamefont{Haggard, Han, Kaminski, and
  Riello}}]{3dblockHHKR}
\bibinfo{author}{\bibfnamefont{H.~M.} \bibnamefont{Haggard}},
  \bibinfo{author}{\bibfnamefont{M.}~\bibnamefont{Han}},
  \bibinfo{author}{\bibfnamefont{W.}~\bibnamefont{Kaminski}}, \bibnamefont{and}
  \bibinfo{author}{\bibfnamefont{A.}~\bibnamefont{Riello}}
  (\bibinfo{year}{2015}{\natexlab{b}}), \eprint{1512.07690}.

\bibitem[{\citenamefont{Han}(2016)}]{hanSUSY}
\bibinfo{author}{\bibfnamefont{M.}~\bibnamefont{Han}}, \bibinfo{journal}{JHEP}
  \textbf{\bibinfo{volume}{01}}, \bibinfo{pages}{065} (\bibinfo{year}{2016}),
  \eprint{1509.00466}.

\bibitem[{\citenamefont{Han and Huang}(2017{\natexlab{a}})}]{Han:2017geu}
\bibinfo{author}{\bibfnamefont{M.}~\bibnamefont{Han}} \bibnamefont{and}
  \bibinfo{author}{\bibfnamefont{Z.}~\bibnamefont{Huang}}
  (\bibinfo{year}{2017}{\natexlab{a}}), \eprint{1702.03285}.

\bibitem[{\citenamefont{Han and Huang}(2017{\natexlab{b}})}]{Han:2016dnt}
\bibinfo{author}{\bibfnamefont{M.}~\bibnamefont{Han}} \bibnamefont{and}
  \bibinfo{author}{\bibfnamefont{Z.}~\bibnamefont{Huang}},
  \bibinfo{journal}{Phys. Rev.} \textbf{\bibinfo{volume}{D95}},
  \bibinfo{pages}{044018} (\bibinfo{year}{2017}{\natexlab{b}}),
  \eprint{1610.01246}.

\bibitem[{\citenamefont{Haggard
  et~al.}(2016{\natexlab{b}})\citenamefont{Haggard, Han, and
  Riello}}]{curvedMink}
\bibinfo{author}{\bibfnamefont{H.~M.} \bibnamefont{Haggard}},
  \bibinfo{author}{\bibfnamefont{M.}~\bibnamefont{Han}}, \bibnamefont{and}
  \bibinfo{author}{\bibfnamefont{A.}~\bibnamefont{Riello}},
  \bibinfo{journal}{Annales Henri Poincare} \textbf{\bibinfo{volume}{17}},
  \bibinfo{pages}{2001} (\bibinfo{year}{2016}{\natexlab{b}}),
  \eprint{1506.03053}.

\bibitem[{\citenamefont{Hayden et~al.}(2016)\citenamefont{Hayden, Nezami, Qi,
  Thomas, Walter, and Yang}}]{Qi1}
\bibinfo{author}{\bibfnamefont{P.}~\bibnamefont{Hayden}},
  \bibinfo{author}{\bibfnamefont{S.}~\bibnamefont{Nezami}},
  \bibinfo{author}{\bibfnamefont{X.-L.} \bibnamefont{Qi}},
  \bibinfo{author}{\bibfnamefont{N.}~\bibnamefont{Thomas}},
  \bibinfo{author}{\bibfnamefont{M.}~\bibnamefont{Walter}}, \bibnamefont{and}
  \bibinfo{author}{\bibfnamefont{Z.}~\bibnamefont{Yang}}
  (\bibinfo{year}{2016}), \eprint{1601.01694}.

\bibitem[{\citenamefont{Pastawski et~al.}(2015)\citenamefont{Pastawski,
  Yoshida, Harlow, and Preskill}}]{Pastawski:2015qua}
\bibinfo{author}{\bibfnamefont{F.}~\bibnamefont{Pastawski}},
  \bibinfo{author}{\bibfnamefont{B.}~\bibnamefont{Yoshida}},
  \bibinfo{author}{\bibfnamefont{D.}~\bibnamefont{Harlow}}, \bibnamefont{and}
  \bibinfo{author}{\bibfnamefont{J.}~\bibnamefont{Preskill}},
  \bibinfo{journal}{JHEP} \textbf{\bibinfo{volume}{06}}, \bibinfo{pages}{149}
  (\bibinfo{year}{2015}), \eprint{1503.06237}.

\bibitem[{\citenamefont{Han and Huang}(2017{\natexlab{c}})}]{Han:2017uco}
\bibinfo{author}{\bibfnamefont{M.}~\bibnamefont{Han}} \bibnamefont{and}
  \bibinfo{author}{\bibfnamefont{S.}~\bibnamefont{Huang}}
  (\bibinfo{year}{2017}{\natexlab{c}}), \eprint{1705.01964}.

\bibitem[{\citenamefont{Delcamp and Dittrich}(2016)}]{Delcamp:2016dqo}
\bibinfo{author}{\bibfnamefont{C.}~\bibnamefont{Delcamp}} \bibnamefont{and}
  \bibinfo{author}{\bibfnamefont{B.}~\bibnamefont{Dittrich}}
  (\bibinfo{year}{2016}), \eprint{1612.04506}.

\bibitem[{\citenamefont{Dittrich et~al.}(2016)\citenamefont{Dittrich,
  Schnetter, Seth, and Steinhaus}}]{Dittrich:2016tys}
\bibinfo{author}{\bibfnamefont{B.}~\bibnamefont{Dittrich}},
  \bibinfo{author}{\bibfnamefont{E.}~\bibnamefont{Schnetter}},
  \bibinfo{author}{\bibfnamefont{C.~J.} \bibnamefont{Seth}}, \bibnamefont{and}
  \bibinfo{author}{\bibfnamefont{S.}~\bibnamefont{Steinhaus}},
  \bibinfo{journal}{Phys. Rev.} \textbf{\bibinfo{volume}{D94}},
  \bibinfo{pages}{124050} (\bibinfo{year}{2016}), \eprint{1609.02429}.

\bibitem[{\citenamefont{Dittrich et~al.}(2014)\citenamefont{Dittrich,
  Martin-Benito, and Steinhaus}}]{Dittrich:2013voa}
\bibinfo{author}{\bibfnamefont{B.}~\bibnamefont{Dittrich}},
  \bibinfo{author}{\bibfnamefont{M.}~\bibnamefont{Martin-Benito}},
  \bibnamefont{and}
  \bibinfo{author}{\bibfnamefont{S.}~\bibnamefont{Steinhaus}},
  \bibinfo{journal}{Phys. Rev.} \textbf{\bibinfo{volume}{D90}},
  \bibinfo{pages}{024058} (\bibinfo{year}{2014}), \eprint{1312.0905}.

\bibitem[{\citenamefont{Bahr et~al.}(2013)\citenamefont{Bahr, Dittrich,
  Hellmann, and Kaminski}}]{Bahr:2012qj}
\bibinfo{author}{\bibfnamefont{B.}~\bibnamefont{Bahr}},
  \bibinfo{author}{\bibfnamefont{B.}~\bibnamefont{Dittrich}},
  \bibinfo{author}{\bibfnamefont{F.}~\bibnamefont{Hellmann}}, \bibnamefont{and}
  \bibinfo{author}{\bibfnamefont{W.}~\bibnamefont{Kaminski}},
  \bibinfo{journal}{Phys. Rev.} \textbf{\bibinfo{volume}{D87}},
  \bibinfo{pages}{044048} (\bibinfo{year}{2013}), \eprint{1208.3388}.

\bibitem[{\citenamefont{Bahr and Steinhaus}(2017)}]{Bahr:2017klw}
\bibinfo{author}{\bibfnamefont{B.}~\bibnamefont{Bahr}} \bibnamefont{and}
  \bibinfo{author}{\bibfnamefont{S.}~\bibnamefont{Steinhaus}}
  (\bibinfo{year}{2017}), \eprint{1701.02311}.

\bibitem[{\citenamefont{Bahr and Steinhaus}(2016)}]{Bahr:2016hwc}
\bibinfo{author}{\bibfnamefont{B.}~\bibnamefont{Bahr}} \bibnamefont{and}
  \bibinfo{author}{\bibfnamefont{S.}~\bibnamefont{Steinhaus}},
  \bibinfo{journal}{Phys. Rev. Lett.} \textbf{\bibinfo{volume}{117}},
  \bibinfo{pages}{141302} (\bibinfo{year}{2016}), \eprint{1605.07649}.

\bibitem[{\citenamefont{Gambini and Pullin}(2009)}]{Gambini:2008as}
\bibinfo{author}{\bibfnamefont{R.}~\bibnamefont{Gambini}} \bibnamefont{and}
  \bibinfo{author}{\bibfnamefont{J.}~\bibnamefont{Pullin}},
  \bibinfo{journal}{Class. Quant. Grav.} \textbf{\bibinfo{volume}{26}},
  \bibinfo{pages}{035002} (\bibinfo{year}{2009}), \eprint{0807.2808}.

\bibitem[{\citenamefont{Oriti et~al.}(2017)\citenamefont{Oriti, Sindoni, and
  Wilson-Ewing}}]{Oriti:2016ueo}
\bibinfo{author}{\bibfnamefont{D.}~\bibnamefont{Oriti}},
  \bibinfo{author}{\bibfnamefont{L.}~\bibnamefont{Sindoni}}, \bibnamefont{and}
  \bibinfo{author}{\bibfnamefont{E.}~\bibnamefont{Wilson-Ewing}},
  \bibinfo{journal}{Class. Quant. Grav.} \textbf{\bibinfo{volume}{34}},
  \bibinfo{pages}{04LT01} (\bibinfo{year}{2017}), \eprint{1602.08271}.

\bibitem[{\citenamefont{Oriti et~al.}(2016)\citenamefont{Oriti, Pranzetti, and
  Sindoni}}]{Oriti:2015rwa}
\bibinfo{author}{\bibfnamefont{D.}~\bibnamefont{Oriti}},
  \bibinfo{author}{\bibfnamefont{D.}~\bibnamefont{Pranzetti}},
  \bibnamefont{and} \bibinfo{author}{\bibfnamefont{L.}~\bibnamefont{Sindoni}},
  \bibinfo{journal}{Phys. Rev. Lett.} \textbf{\bibinfo{volume}{116}},
  \bibinfo{pages}{211301} (\bibinfo{year}{2016}), \eprint{1510.06991}.

\bibitem[{\citenamefont{Oriti}(2016)}]{Oriti:2016acw}
\bibinfo{author}{\bibfnamefont{D.}~\bibnamefont{Oriti}} (\bibinfo{year}{2016}),
  \eprint{1612.09521}.

\end{thebibliography}


\end{document}